\documentclass[aps,preprint,longbibliography]{revtex4-2}
\usepackage{bm,amsfonts,graphicx,cancel}
\usepackage{latexsym,amssymb,amsmath,color}
\usepackage[utf8]{inputenc}
\usepackage{caption}
\usepackage{subcaption}
\usepackage{ulem}
\usepackage{amsmath}
\usepackage{multirow}
\usepackage{xcolor}
\usepackage{tikz}

\begin{document}

\title{Far field of turbulent spots}
\author{Pavan V. Kashyap$^1$}  %
\author{Yohann Duguet$^1$}
\author{Matthew Chantry$^2$}
\affiliation{$^1$LIMSI-CNRS, UPR 3251, Universit\'e Paris-Saclay, 91405 Orsay, France%
\\
$^2$Atmospheric, Oceanic and Planetary Physics, University of Oxford, Oxford, United Kingdom}

\date{May 2020}

\begin{abstract}
The proliferation of turbulence in subcritical wall-bounded shear flows involves spatially localised coherent structures. Turbulent spots correspond to finite-time nonlinear responses to pointwise disturbances and are regarded as seeds of turbulence during transition. The rapid spatial decay of the turbulent fluctuations away from a spot is accompanied by large-scale flows with a robust structuration. The far field velocity field of these spots is investigated numerically using spectral methods in large domains in four different flow scenarios (plane Couette, plane Poiseuille, Couette-Poiseuille and a sinusoidal shear flow). At odds with former expectations, the planar components of the velocity field decay algebraically. These decay exponents depend only on the symmetries of the system, which here depend on the presence of an applied gradient, and not on the Reynolds number. This suggests an effective two-dimensional multipolar expansion for the far field, dominated by a quadrupolar flow component or, for asymmetric flow fields, by a dipolar flow component.
\end{abstract}

\maketitle

\section{Introduction} \label{S1}

Transition to turbulence in wall-bounded shear flows requires usually finite-amplitude initial disturbances, and the transition process is intrinsically spatiotemporal \citep{Tuckerman2017arfm}. A zoo of different large-scale coherent structures surrounded by quasi-laminar flow have been observed in early experiments and simulations: approximately circular spots \citep{emmons1951laminar}, oblique stripes \citep{Coles1965transition,Prigent2002large} and labyrinths \citep{Duguet2010formation, chantry2016turbulent}. These structures display complex temporal dynamics on long time scales including symmetry breakings \citep{Duguet2013oblique}, spontaneous relaminarisation \citep{Bottin1998statistical} and self-replication \citep{paranjape2019thesis}.  Here we study turbulent spots, which we define as nonlinear responses of the system, in finite time, to a spatially localised disturbance. These coherent structures are not steady states of the system but transients : each spot either decays or proliferates in time. Spots have initially been described by \citet{emmons1951laminar} following water table experiments. They have later been identified in early experiments of channel flow \cite{Carlson1982flow,Alavyoon1986turbulent,HenningsonKim1991spots}, in plane Couette flow \cite{tillmark1992experiments,dauchot1994finite,Dauchot1995finite,couliou2015large,couliou2016spreading,couliou2017growth}, and are frequently found in the bypass regime of boundary layer flows \citep{dhawan1958some,matsubara2001disturbance, kreilos2016bypass, khapko2016edge}. Spots can be considered as the two-dimensional equivalent of the turbulent 'flashes' initially described by O. Reynolds in cylindrical pipe flow, nowadays called 'puffs' \cite{Reynolds1883xxix,Wygnanski1973transition}.
Apart from their synthetic generation, spots are also found at the lower end of the transitional regime of plane Couette flow \citep{Duguet2010formation,manneville2011decay,desouza2020transient}. The dynamics at onset involves long-range correlations and is thought to correspond to a universal continuous phase transition \cite{Lemoult2016directed,Chantry2017universal}. The way these individual protagonists interact together is hence crucial for better theoretical understanding and modelling of these complex regimes. In particular, the way two remote turbulent spots interact together as a function of their distance is key information for the understanding of the self-organisation of a larger assembly of spots. This interaction problem has been addressed recently for puffs in cylindrical pipe flow \citep{barkley2011simplifying,mukund2018critical,lemoult2020statistical}. Numerical estimations of the decay of the velocity field outside a localised puff suggests an exponential decay for all components as a function of the streamwise distance \citep{samanta2011experimental, Ritter2016emergence}. This essentially suggests short-range interactions, a crucial ingredient to test the directed percolation hypothesis \citep{goldenfeld2017turbulence}. 
As for planar flows however, the situation is less clear. Whereas the interaction between adjacent oblique turbulent stripes also appears to be exponential \citep{paranjape2020oblique,gome2020statistical}, re-examination of numerical simulations (in e.g. \cite{Duguet2010formation}) suggest that turbulent spots display a slower spatial decay. Earlier analysis of a sinusoidal shear flow \citep{schumacher2001evolution} suggests exponential decay, although only the decay of the wall-normal velocity is reported. A more recent analysis of a doubly localized steady solution of plane Couette flow \cite{Brand2014doubly} suggests exponential decay for the streamwise velocity component, justified by an analytical ansatz. More recently, an analytical study of plane Couette flow has suggested in turn algebraic decay of the planar velocity components with a decay exponent of +3 \citep{zhe2020}. On one hand no analytical approach is free from the artificial simplifications aimed at making the problem solvable. On the other hand the main technical obstacles for confirming or overturning such scalings from data are finite-size effects: distinguishing unambiguously between exponential and algebraic decay requires several decades of spatial extent. Numerical simulations usually rely on periodic boundary conditions, which are poorly compatible with smooth spatial localisation. As a consequence huge computational domains are needed, which has not been attempted as of now.

Experimentally the problem is even steeper: accurate measurements of velocity tails in a noisy environment are technically very difficult because of the very low amplitudes involved. The present numerical investigation fills the gap by showing quantitative evidence that spots are surrounded by large-scale flows decaying algebraically, and that the associated exponent is independent of the Reynolds number. Four examples of wall-bounded shear flows have been chosen : plane Couette flow (pCf), plane Poiseuille flow (pPf), Couette-Poiseuille flow (CPf) and the sinusoidal shear flow of Ref. \cite{schumacher2001evolution} also called sometimes Waleffe flow (Wf) \cite{chantry2016turbulent}. All of them have a linearly stable laminar profile for the range of values of Reynolds number explored.\\
The paper is structured as follows : Section~\ref{S2} is devoted to the numerical methodology and an early evidence for algebraic decay. A more detailed estimation of the decay exponents is carried out in Section~\ref{S3}. The results are discussed in Section~\ref{S4}. \\

\section{Methodology} \label{S2}

\subsection{Flow scenarios \label{S2.1}}

We begin by defining the conventions used for the four flow scenarios under scrutiny. For the sake of generality, four different flow scenarios have been chosen, exploring various combinations of flow symmetries, pressure gradient forcing and differing boundary conditions. Due to computational constraints these also feature different degrees of numerical resolution, ranging from poor to excellent. \\

\begin{figure}[h]
    \centering
    \begin{subfigure}[b]{0.45\textwidth}
        \centering
         \includegraphics[scale=1]{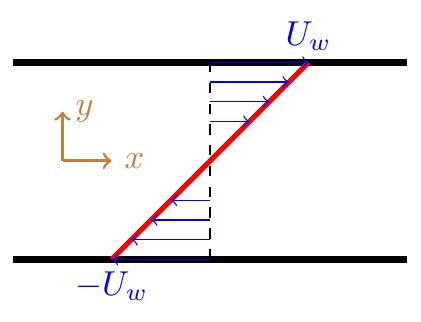}
        \caption{}
        \label{pcf}
    \end{subfigure}
    \hfill
    \begin{subfigure}[b]{0.45\textwidth}
        \centering
        \includegraphics[scale=1]{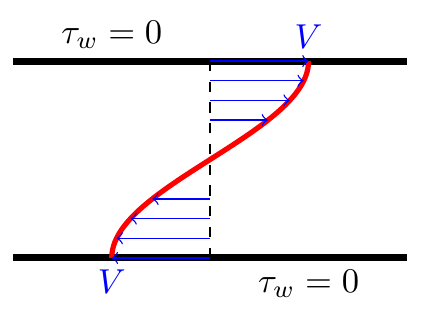}
        \caption{}
        \label{Wf}
    \end{subfigure}
    \hfill
    \begin{subfigure}[b]{0.45\textwidth}
        \centering
        \includegraphics[scale=1]{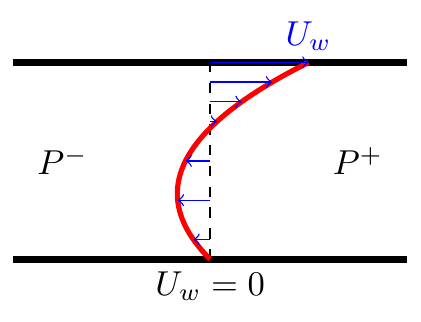}
        \caption{}
        \label{cPf}
    \end{subfigure}
    \hfill
    \begin{subfigure}[b]{0.45\textwidth}
        \centering
        \includegraphics[scale=1]{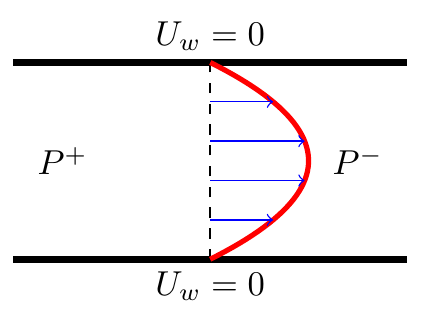}
        \caption{}
        \label{ppf}
    \end{subfigure}
        \caption{Laminar base flow profiles for (a) plane Couette flow (pCf) (b) Waleffe flow (Wf)  (c) Couette-Poiseuille flow (CPf) and (d) plane Poiseuille flow (pPf). $\tau_w$ stands for the wall shear stress $\rho \nu \partial u /\partial y$, $U_w$ for the velocity at the wall, while the labels $P^+$ and $P^-$ indicate an applied pressure gradient.}
        \label{flow}
\end{figure}

These four flows are governed by the same incompressible Navier-Stokes equations
\begin{align}
\frac{\partial \bm{u}}{\partial t} + (\bm{u} \cdot  \nabla) \bm{u} &= -\frac{1}{\rho}\nabla p + \nu \nabla^2 \bm{u} + {\bm f} \label{ns} \\
\nabla \cdot \bm{u} &= 0 \label{mc}
\end{align}

where $\bm{u}=(u_x,u_y,u_z)$ is the dimensional velocity field, $\rho$ the density of the fluid and $\nu$ its kinematic viscosity. Only the boundary conditions and the choice for the forcing term ${\bm f}$ distinguish one flow from another one. In all scenarios the flow occurs between two walls located respectively at $y=-h$ (lower wall) and $y=+h$ (upper wall). The streamwise and spanwise coordinates are respectively called $x$ and $z$.\\

The laminar flow for plane Couette flow (pCf), Waleffe flow (Wf) and Couette-Poiseuille flow (cPf) have zero mass flux. In pCf both walls at $y=\pm h$ move with opposite velocities $\pm U_w$, with $U_w>0$. In cPf only the upper wall at $y=h$ moves with velocity $U_w$ while the other is stationary. A fixed streamwise pressure gradient is imposed such that the laminar flow rate is exactly zero (as in Ref. \cite{Klotz2017couette}). Wf is forced by a $y$-dependent body force $f_x \sim \sin\left(\pi y/2h\right)$ and no mass flux is permitted in any direction. Plane Poiseuille flow (pPf) is driven by imposing a mean streamwise pressure gradient with stationary walls. In pCf, cPf and pPf scenarios the mean spanwise pressure gradient is set to zero. \\

The different flow scenarios have different non-dimensionalisations. The control parameter in each case is a Reynolds number $Re=UL/\nu$ based on a lengthscale $L$ and a velocity scale $U$. The lengthscale $L$ is always the half-height of the channel $h$. The velocity scale $U$ depends however on the flow case under consideration. The notation for the Reynolds number of each flow case follows the conventions in Table~\ref{Re}. From here on only non-dimensional velocities $\bm{U}=\bm{u}/U$ will be used, with components $U_x$, $U_y$ and $U_z$. The nondimensional timescale is defined as $t=h/U$.

\begin{table}[h]
\begin{tabular}{|c|c|c|}
\hline
Flow Case & Reynolds number & Velocity scale\\
\hline \hline
pCf & $ Re = \frac{U_w \, h}{\nu} $ & $U_w$ - Wall velocity \\
\hline
pPf & $ Re_{cl} = \frac{u_{cl} \, h}{\nu} $ & $U_{cl}$ - Laminar center line velocity \\
\hline
cPf & $Re_w = \frac{U_w \, h}{\nu} $ &  $U_w$ - Velocity of the moving wall\\
\hline 
Wf & $Re_V = \frac{V \, h}{\nu}$ & $V$ - velocity at wall \citep{chantry2016turbulent}\\
\hline
\end{tabular}
\caption{Reynolds number definition and velocity scale for each flow case.}
\label{Re}
\end{table}

The laminar base flow solutions for each flow case are, in non-dimensional form,
\begin{align}
\textrm{pCf}  :  U_{L}(y_{*}) &= y_{*} \label{pcfeq}\\
\textrm{pPf}  :  U_{L}(y_{*}) &= (1-y_{*}^2) \label{ppfeq}\\
\textrm{cPf}  :  U_{L}(y_{*}) &= \frac{3}{4}(y^2_{*}-1)+\frac{1}{2}(y_*+1) \label{cpfeq}\\
\textrm{Wf}   :  U_{L}(y_{*}) &= \sin(\beta y_*) \label{wfeq}
\end{align}

where $y_*=y/h$. For simplicity the subscript `*' will be dropped for the rest of the paper.

\subsection{Numerical methods} \label{S2.2}

Direct numerical simulation of the unsteady Navier-Stokes equations (\ref{ns}-\ref{mc}) is performed using the efficient, open-source, parallel spectral code \emph{ChannelFlow2.0} \cite{Channelflow} for pCf, CPf and pPf. This algorithm relies on a Fourier-Chebyshev-Fourier pseudo-spectral discretization of the velocity field expressed in primitive variables. The time-stepping algorithm is a 3$^{rd}$ order semi-implicit backwards differentiation scheme. Boundary conditions in the $y$-direction are no slip and are imposed via the influence matrix method with Chebyshev--tau correction. The $(x,z)$ boundary conditions are periodic.

The Waleffe Flow (Wf) simulations correspond to the configuration called the model Waleffe Flow \cite{chantry2016turbulent} by Chantry \emph{et al.}. The simulation makes use of Fourier-Fourier-Fourier pseudo-spectral discretization of the equations in mean-poloidal-toroidal formulation, advanced using a combination of backward Euler and Adams-Bashforth for the linear and nonlinear terms respectively. This flow has free-slip boundary conditions in the $y$-direction and periodic boundary conditions in the $(x,z)$-directions. Further numerical details can be found in Ref~\cite{chantry2016turbulent}.\\

For all the flow scenarios the normalised domain size is $(L_x,2,L_z)$ and we keep $L_x = L_z$ throughout this study. The spatial discretization is specified by the integer triplet $(J,K,L)$ or by the corresponding physical grid given by the triplet $(N_x,N_y,N_z)$ = $(2J +1,L ,2K+1)$. Nonlinear terms are computed using the 3/2 dealiasing rule. pCf at $Re=400$ is our reference case used to tailor the working parameters. The resolution used for all the simulations of pCf, cPf and pPf is $N_x/L_x,N_y,N_z/L_z=4,33,8$ with an increase to $N_y=65$ for higher values of \emph{Re}. The effectiveness of this resolution is discussed in section~\ref{S2.4}. The Wf simulations were performed at reduced resolution using $N_x/L_x,N_y,N_z/L_z=3.2,4,3.2$ just as in Ref~\cite{chantry2016turbulent}.

The initial disturbance to the laminar flow consists in all scenarios of a localized disturbance to the laminar field. The same disturbance as in  \cite{lundbladh1991direct} and \cite{Duguet2009localized} has been considered, rotated by an arbitrary angle in the $(x,z)$ plane in order to discard any symmetry. A standard discrete Fourier transform is then applied to this disturbance in order to generate an initial condition for the spectral code. The amplitude of this initial disturbance does not matter as long as it is strong enough to trigger transition. Note that the initial condition does not contain any large-scale flow, a transient time is hence necessary in the simulations for the emergence of that flow and the build-up of the tails. An order of magnitude for this transient for the domain size $L_x=L_z=1280$ is $t \approx 50$.

\subsection{Topology of large-scale flows} \label{S2.3}

We detail here the topology of the flow in the far field. It is well known that the flow is not strictly identical to the laminar solution outside the ``turbulent'' zone where Reynolds stresses are not negligible. Just outside this zone, non-turbulent large-scale flows are present and decay away from the spot \citep{lundbladh1991direct,li1989wave,henningson1991turbulent}. Characterizing the far field of the turbulent spots is equivalent to probing the decay of the large-scale velocity field generated by the spot. A scale separation hypothesis is often used in order to consider the large-scale flow steady over timescales shorter than the timescale for the growth of the spot. Such a hypothesis is valid when the spot grows slowly enough, i.e. close to onset \cite{Dauchot1995finite,couliou2016spreading}.

\begin{figure}[h]
     \centering
     \begin{subfigure}[b]{0.48\textwidth}
         \centering
         \includegraphics[width=\textwidth]{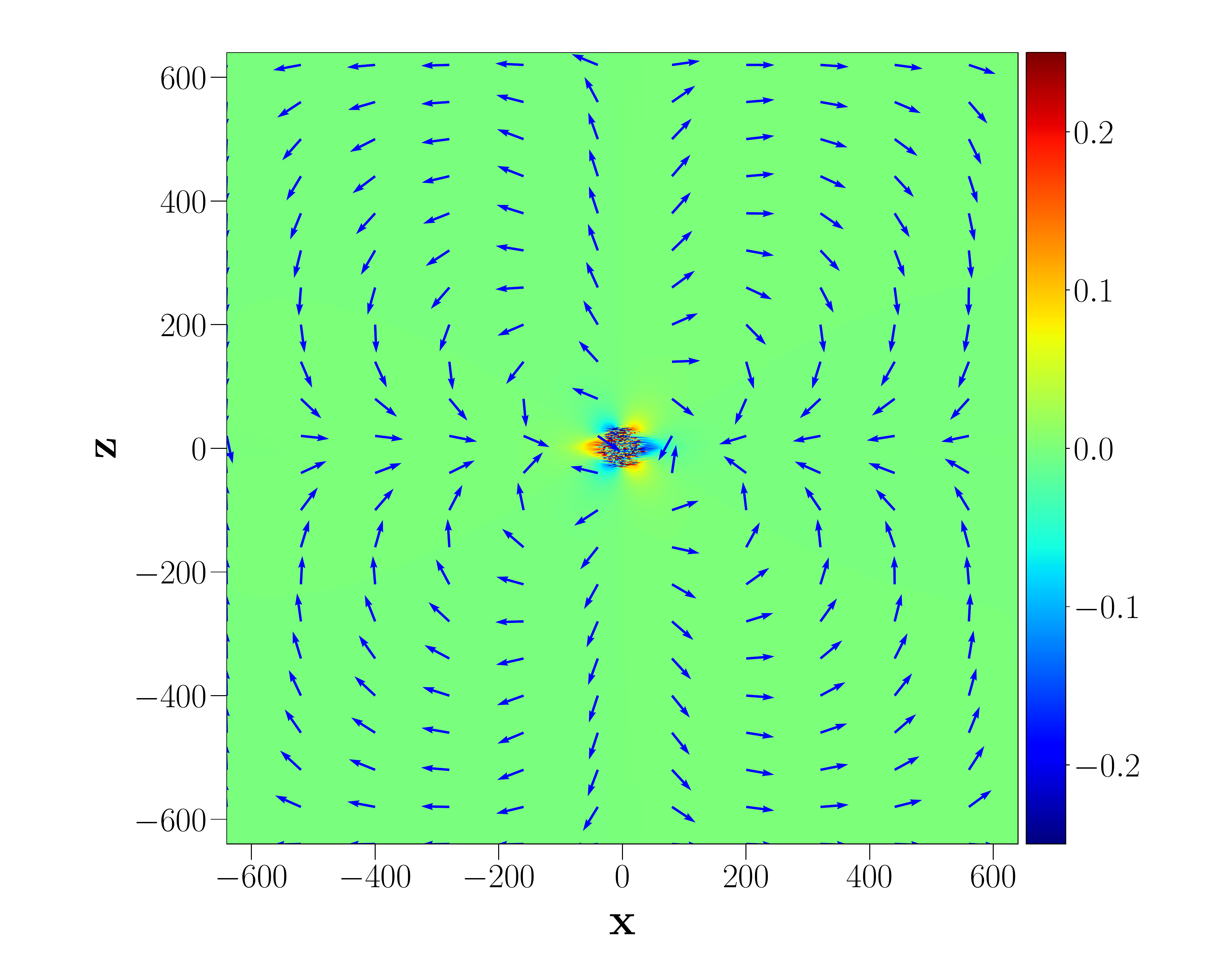}
         \caption{}
         \label{pcf_spot}
     \end{subfigure}
     \hfill
     \begin{subfigure}[b]{0.48\textwidth}
         \centering
         \includegraphics[width=\textwidth]{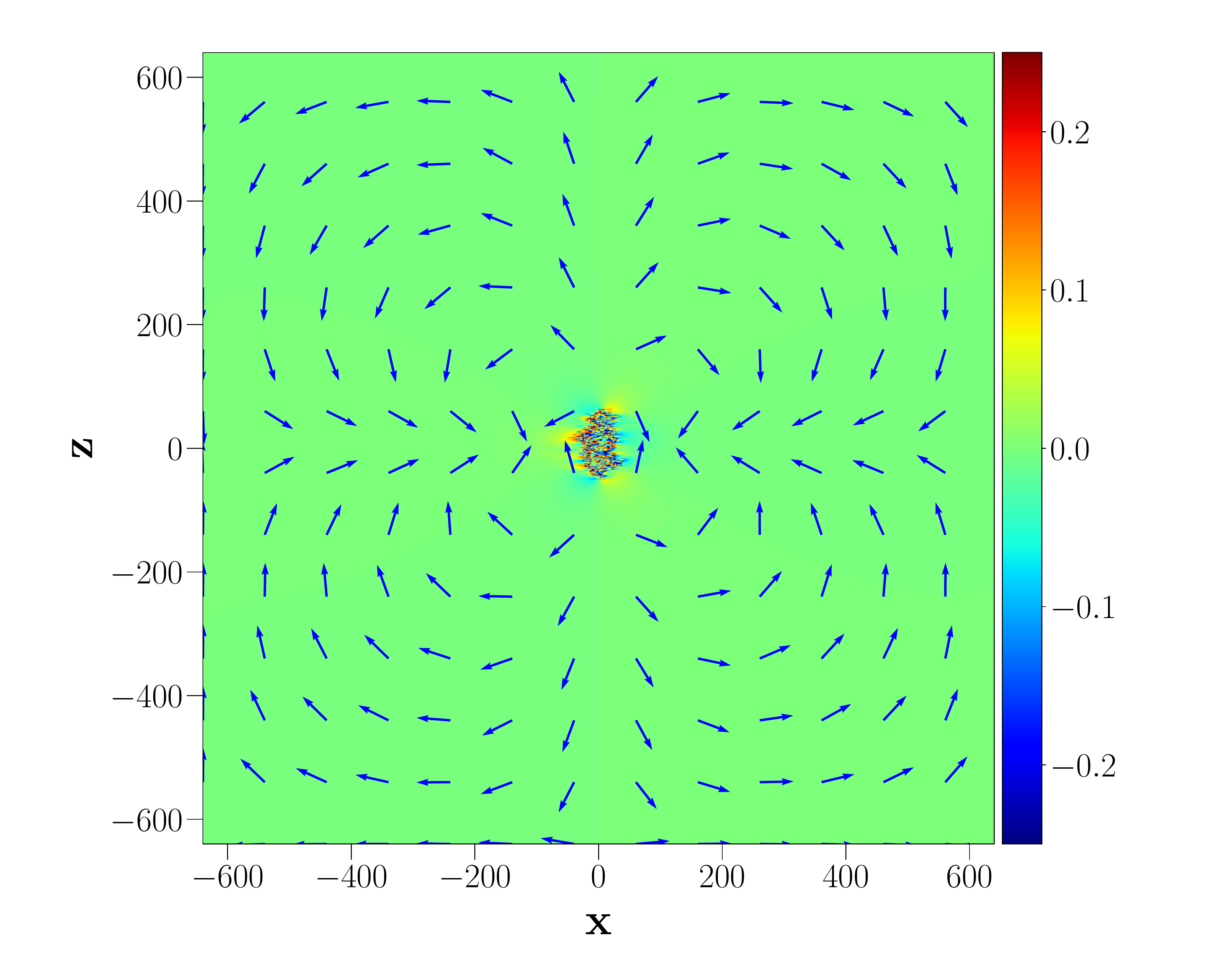}
         \caption{}
         \label{wf_spot}
     \end{subfigure}
     \hfill
     \begin{subfigure}[b]{0.48\textwidth}
         \centering
         \includegraphics[width=\textwidth]{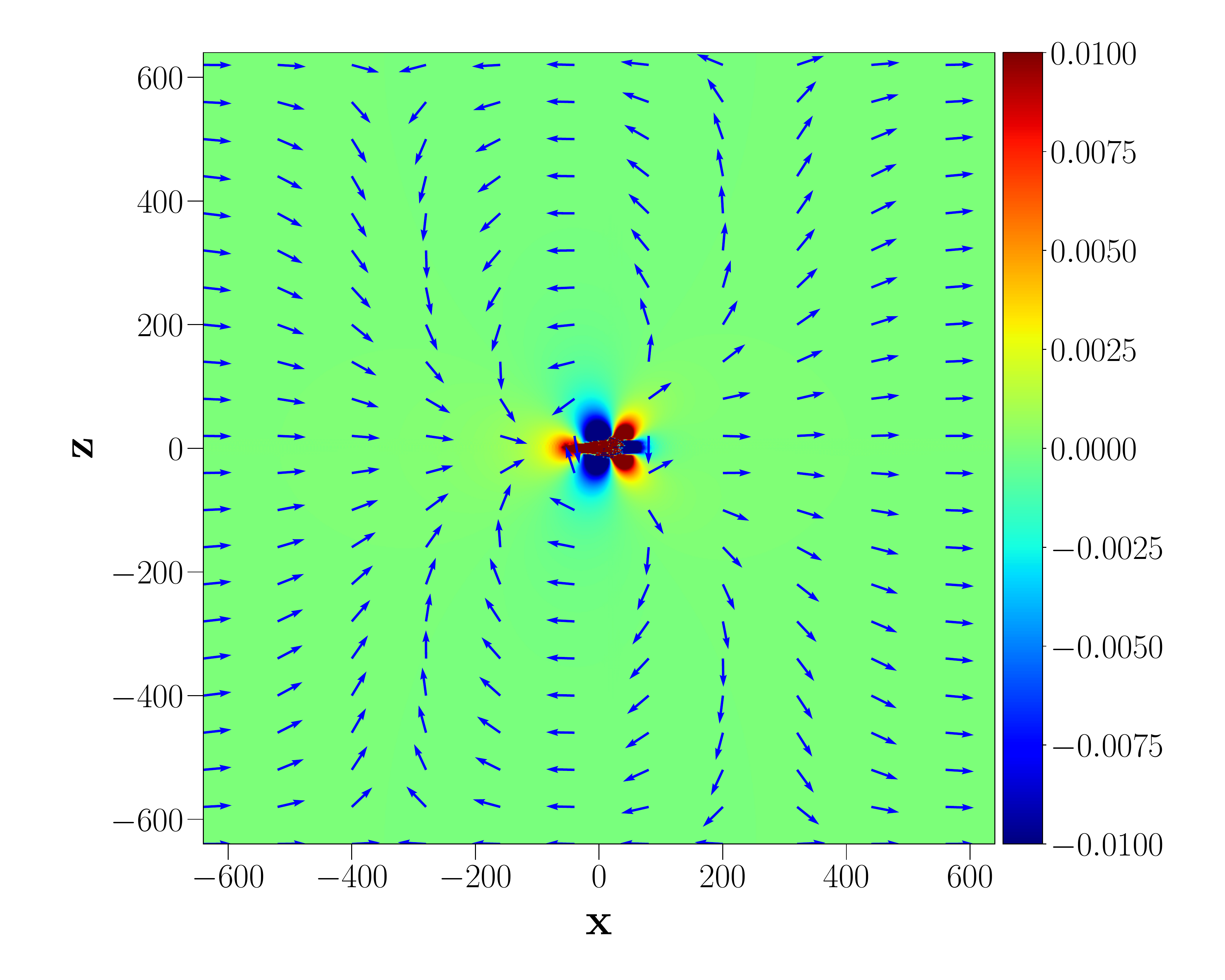}
         \caption{}
         \label{cpf_spot}
     \end{subfigure}
     \hfill
     \begin{subfigure}[b]{0.48\textwidth}
         \centering
         \includegraphics[width=\textwidth]{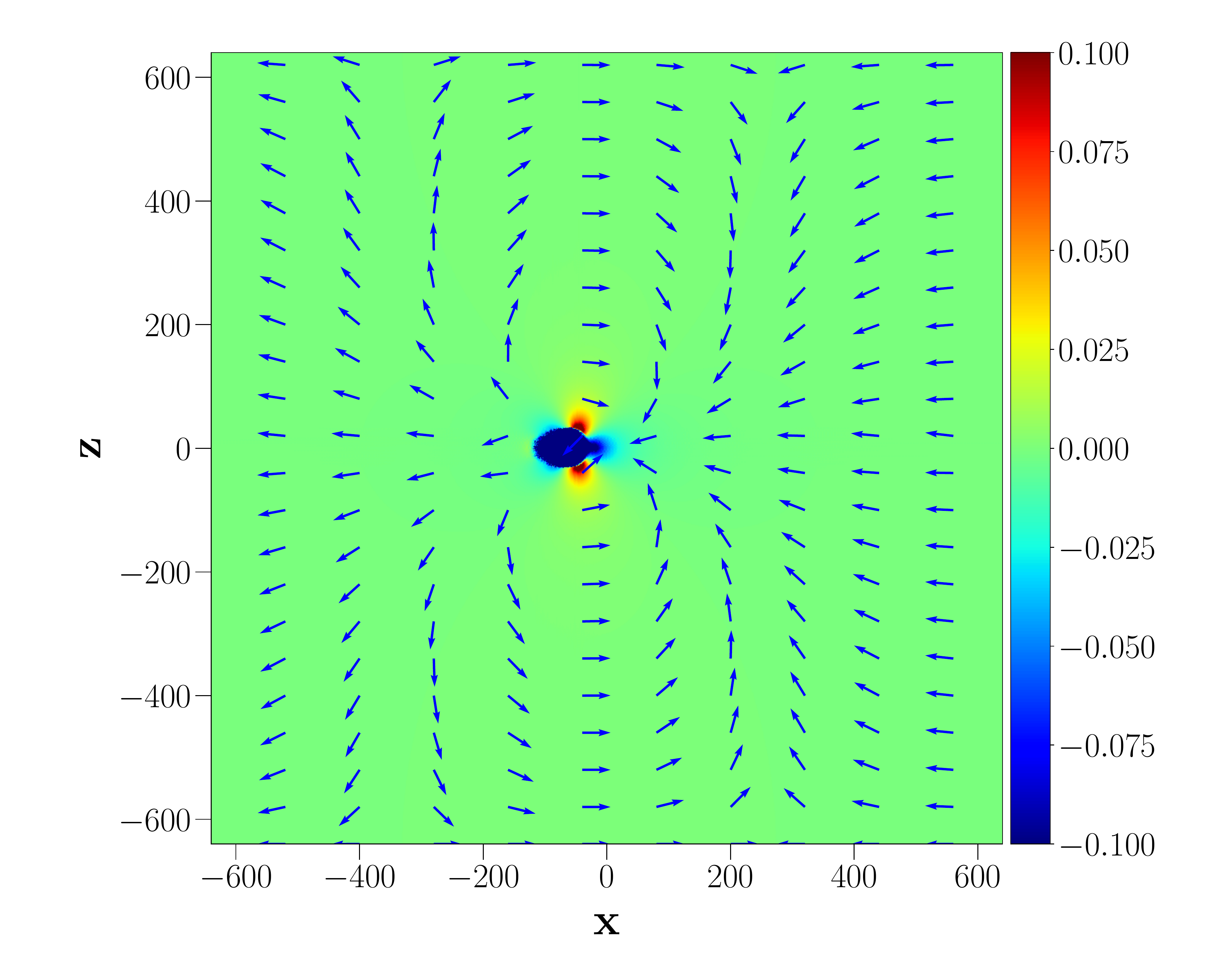}
         \caption{}
         \label{ppf_spot}
     \end{subfigure}
        \caption{ Turbulent spots in the different flow scenarios visualized by isocontours of the deviation from the laminar field $\left( U_x - U_L \right)$ at $y=0$ with the associated large-scale flow ($\bar{U}_x,\bar{U}_z$)depicted by their unit vectors : (a) pCf at $Re=400$ and $T-300$ (b) Wf at $Re_V=83.5$ and $T=250$ (c) CPf at $Re_W=700$ and $T=200$ (d) pPf at $Re_{cl}=3200$ at $T=200$.}
        \label{spots}
\end{figure}

In the case of plane Couette flow, the origin of these large-scale flows is the mismatch between different flow rates across the interface of the turbulent zone. This mismatch leads to long wake regions called `overhangs' \cite{lundbladh1991direct,lagha2007modeling,Duguet2013oblique}. It generates invariably a flow towards the interior of the spot along the $x$ axis. By virtue of incompressibility this streamwise flow is compensated by a flow outside the spot in the spanwise direction. Away from these two axis, the large-flow induces large vortices and the main structure resembles a quadrupole in the $(x,z)$-plane. The situation is similar in Wf, as shown in Ref. \cite{chantry2016turbulent}. In Ref. \cite{zhe2020} inverted quadrupoles were also mathematically predicted for exotic forcing parameters. Such inverted quadrupoles have not been observed in any of our simulations, and they have no relevance to the far field anyway. 
As for the other flows including pressure gradients, spots with a quadrupolar-looking large-scale flow were also reported in recent experiments of plane Poiseuille flow \cite{Lemoult2014turbulent} and Couette-Poiseuille flow \cite{klotz2017transition,klotz2017experiments}, however the effect of the confinement by the side walls has not been investigated. From our results in fig. \ref{ppf_spot}, the large-scale flow inside the turbulent region is equivalent to the turbulent mean flow correction, counteracting the laminar mean flow. By continuity, two recirculations appear one each side of the streamwise axis, akin to a dipole with a dipolar moment aligned with the spanwise direction. This suggests that spots in flows with a pressure gradient are expected to be dominated by a dipolar flow field rather than a quadrupolar one. For the case of CPf, the picture is the same except that the pressure gradient is oriented opposite to that in pPf. Consequently the dipolar moment of CPf also points into the direction opposite to that of pPf. \\

The structure of these large-scale flows is three-dimensional, however the present analysis is based on a two-dimensional (planar) description. We use the $y$-averaged velocity field $\bar{{\bm U}}$ defined by its components
\begin{equation}
\bar{U}_x=\int \left(U_x - U_{L}\right) \, dy \quad , \quad \bar{U}_y = \int U_y \, dy \quad ,  \quad \bar{U}_z=\int U_z \, dy. \label{lsfeq}
\end{equation}

The two-dimensional field $(\bar{U}_x,\bar{U}_z)$ is divergence-free with respect to the two variables $x$ and $z$ \cite{Duguet2013oblique}. An example of such turbulent spot is shown in Figure~\ref{pcf_spot} along with the planar large-scale flow $(\bar{U}_x,\bar{U}_z)$, for the reference case of pCf at $Re=400$, evaluated after 300 non-dimensional time units. In this figure, the velocity vectors have been normalised in order to highlight the far field. The turbulent spot for pCf indeed has the expected quadrupolar structure reported in former works. The large-scale flow for Wf, displayed in fig.\ref{wf_spot}, has a similar structure. The large-scale flow associated with CPf and pPf, respectively in figs \ref{cpf_spot} and \ref{ppf_spot}, do not show any clear quadrupolar structure. They have a structure very similar to that of a dipolar flow with a dipolar moment oriented along the $z$ direction. This distinction will be further analysed in Section~\ref{S3}. \\

\subsection{Evidence for algebraic decay} \label{S2.4}

In order to accurately capture the far field behaviour in the absence of finite-size effects, large enough numerical domains are required. The quantitative influence of $L_x$ and $L_z$ (assumed here equal for a square domain) is assessed for the reference case of pCf for $Re=400$. 
\begin{figure}[h]
\centering
	\begin{subfigure}[b]{0.49\textwidth}
	\centering
	\includegraphics[width=\textwidth]{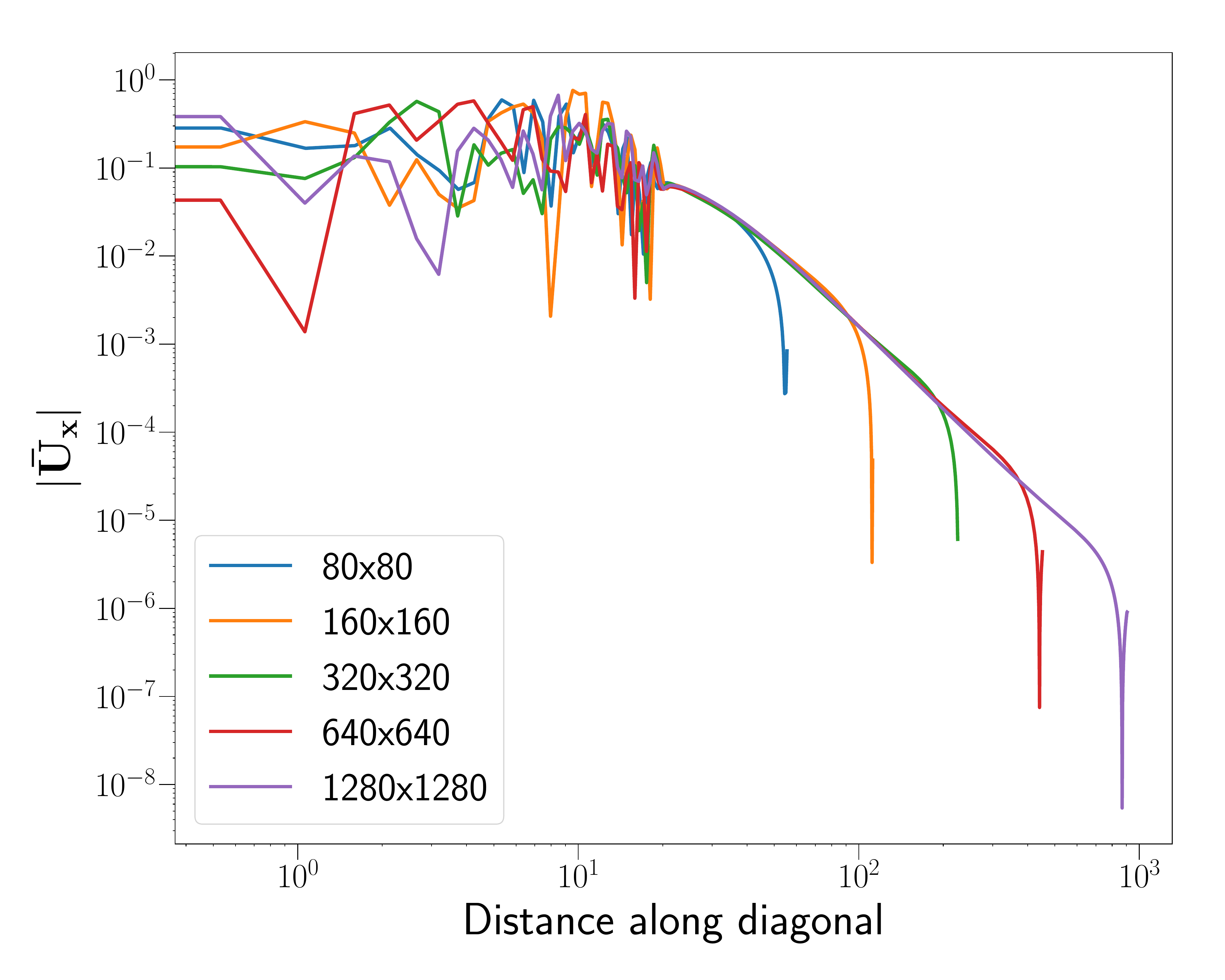}
	\caption{}
	\end{subfigure}
	\begin{subfigure}[b]{0.49\textwidth}
	\centering
	\includegraphics[width=\textwidth]{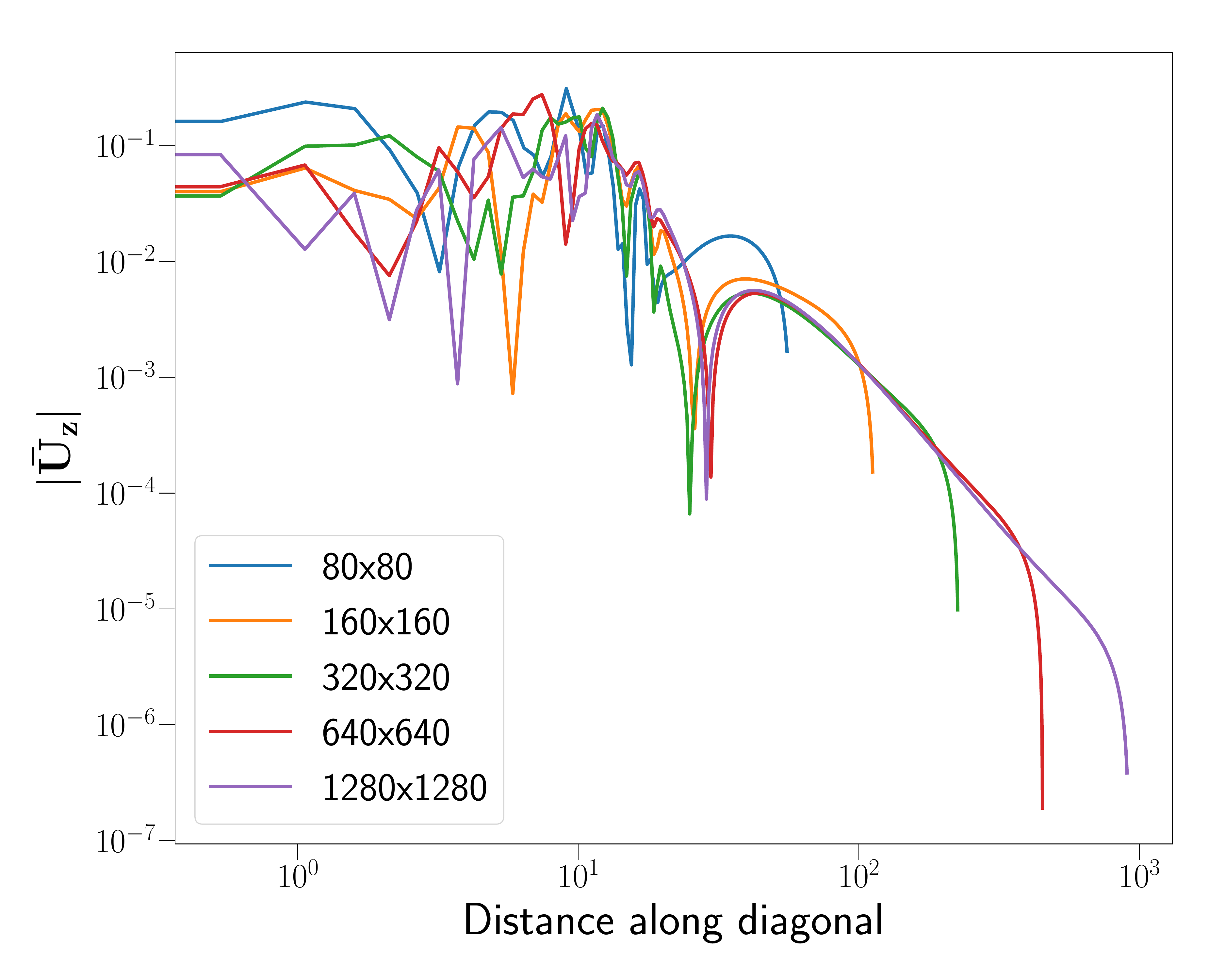}
	\caption{}
	\end{subfigure}
	\caption{Decay of the velocity tails along the diagonal $x=z$ of the domain for different domain sizes (shown in legend as $L_x \times L_z$) for pCF at $Re=400$ and $T=150$ (a) Streamwise component $\bar{U}_x$ (b) Spanwise component $\bar{U}_z$} 
	\label{boxsize}
\end{figure}
Figure~\ref{boxsize} shows the decay of $\bar{U}_x$ and $\bar{U}_z$ monitored along the diagonal line $x$=$z$ (displayed for $x>0$ only). The choice for the diagonal allows for a wider spatial extent, however the results have been checked to be independent of the angle in the $(x,z)$-plane. \\

\begin{figure}
\centering
\includegraphics[scale=0.32]{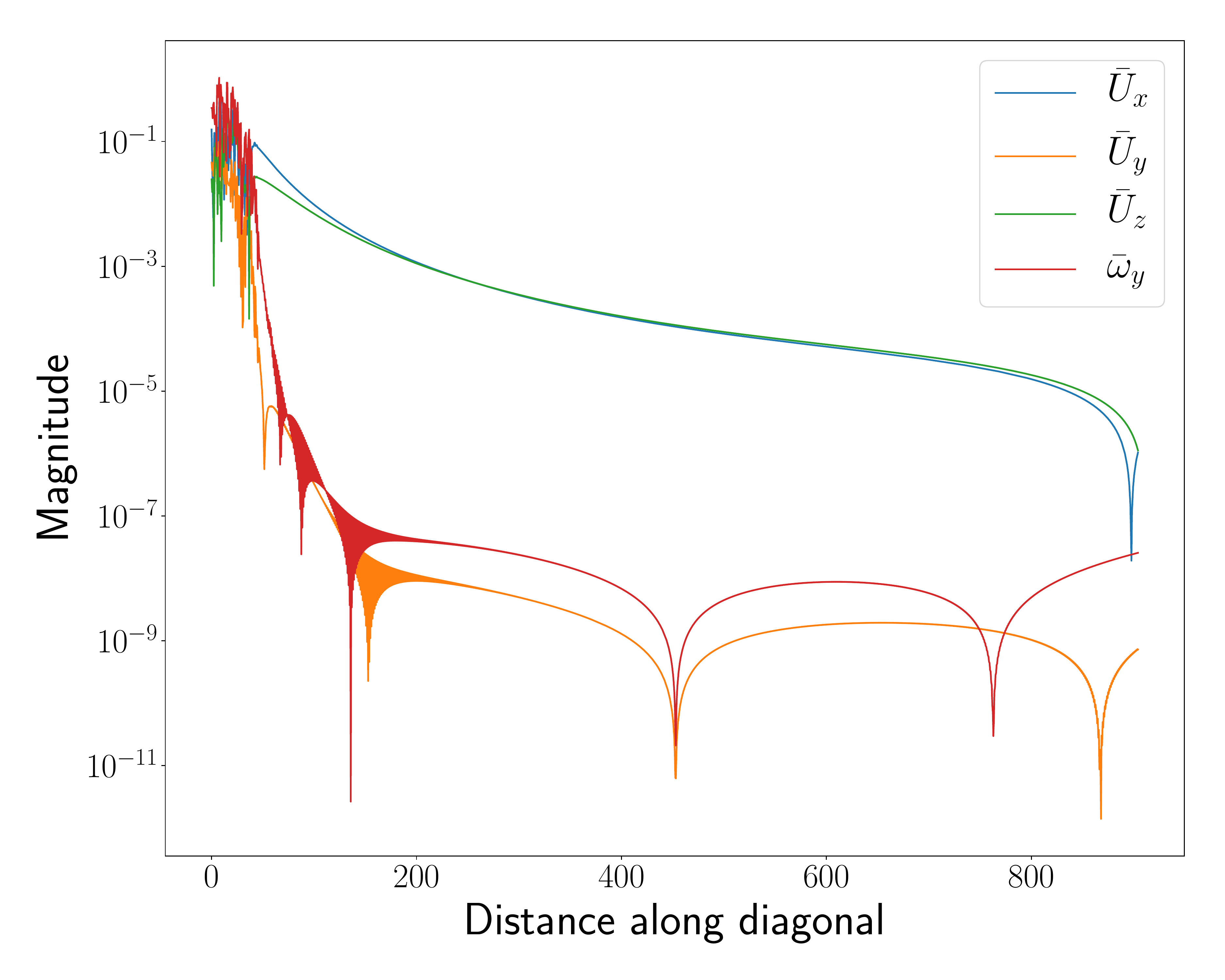}
\caption{Variation of wall $y$-averaged variable $\bar{U}_x, \, \bar{U}_y, \, \bar{U}_z$ and $\bar{\omega}_y$ along the diagonal of the domain $x=z$ for pCf at $Re=400$ and $T=300$.}
\label{Vy}
\end{figure}

For $L_x \lesssim 200$ the decay of $\bar{U}_x$ and $\bar{U}_z$ can be described as exponential. However, increasing $L_x$ up to 1280 increasingly reveals algebraic decay. This algebraic decay is evident in log-log coordinates over one and a half decade. This directly suggests that, in the far field limit, the decay of both $\bar{U}_x$ and $\bar{U}_z$ is dominated by an algebraic flow component. The decay exponent of both $\bar{U}_x$ and $\bar{U}_z$ is close to 3 as will be evaluated carefully in the next section. Not all fields decay algebraically, however. The $y$-integrated wall-normal velocity $\bar{U}_y$ decays faster than exponentially in absolute value, independently of the domain size once it is large enough. Interestingly, although it derives from $\bar{U}_x$ and $\bar{U}_z$ which decay algebraically, the $y$-integrated wall-normal vorticity $\bar{\omega}_y=\partial_z \bar{U}_x-\partial_x \bar{U}_z$ decays faster than an exponential. Both quantities are displayed in Figure~\ref{Vy}. This suggests that, to a decent approximation, the $y$-integrated large-scale flow is irrotational outside the core of the spot. These results match the analytical predictions in Ref. \cite{zhe2020} despite the strong hypotheses formulated in that work, notably the dubious but inevitable trade-off in the boundary conditions. In the next section we analyse the generality of the algebraic decay by: i) testing the dependence on the numerical resolution, ii) testing different flow scenarios with different boundary conditions and iii) conducting a parametric study on the dependence of \emph{Re}.\\

\begin{figure}
	\begin{minipage}{.55\textwidth}
	\centering
	\includegraphics[width=\textwidth]{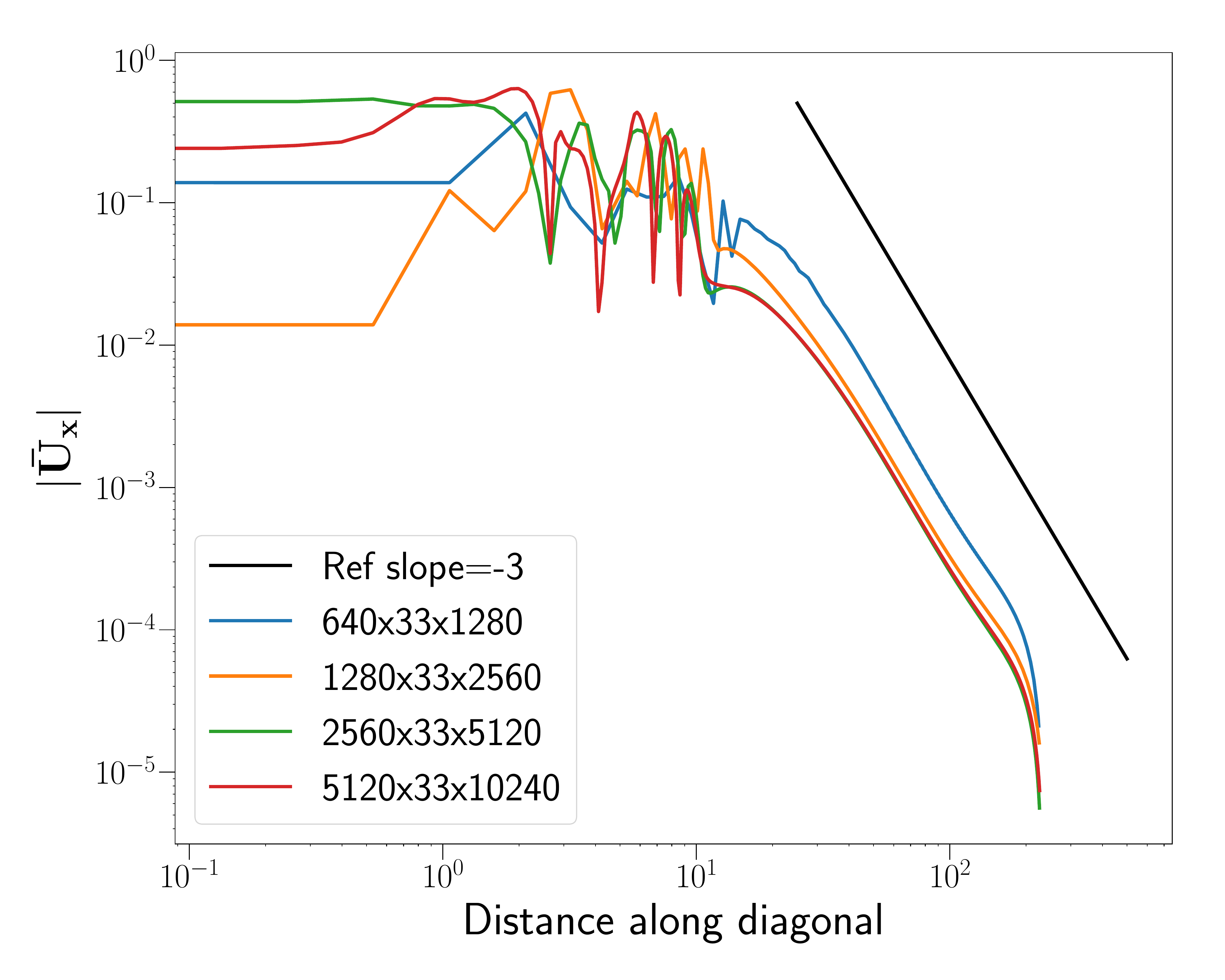}
	\captionof{figure}{Decay of $\bar{U}_x$ along the diagonal $x=z$ of the domain $L_x=L_z=320$, for increasing spectral resolutions (shown in legend as $N_x \times N_y \times N_z$) for pCf at $Re=400$ and $T=80$.}
	\label{rescheck}
	\end{minipage}\hfill
	\begin{minipage}{.45\textwidth}
    \centering 
	\begin{tabular}{|c|c|c|c|c|c|}
		\hline
		$L_x$ & $L_z$ & $N_x$ & $N_y$ & $N_z$ & Exponent($U_x$) \\
		\hline \hline
		\multirow{4}{*}{320} & \multirow{4}{*}{320} & 640 & 33 & 1280 & $ +2.95 \pm 0.08$ \\
		\cline{3-6}
		 &  & 1280 & 33 & 2560 & $ +3.07 \pm 0.025 $ \\ 
		\cline{3-6}
		 &  & 2560 & 33 & 5120 & $ +3.08 \pm 0.01$ \\ 
		\cline{3-6}
		 &  & 5120 & 33 & 10240 & $ +3.07 \pm 0.04$ \\ 
		\hline
		\end{tabular}
	\captionof{table}{Decay exponents computed for different resolutions in the domain $L_x=L_z=320$ for $\bar{U}_x$ of pCf at $Re=400$ and $T=80$}
	\label{expcheck}
	\end{minipage}
\end{figure}

The results of the parametric study in Figure~\ref{boxsize} indicate that for the reference case of pCf at $Re=400$, a square numerical domain with $L_x=L_z=1280$ and a resolution of $N_x,N_y,N_z=5120,33,10240$ conclusively capture the algebraic decay. The adequacy of this resolution was tested on a smaller box of $L_x=L_z=320$. The resolution was varied from $N_x/L_x=2$ to $16$ by factors of two. The algebraic decay of $\bar{U}_x$ along the diagonal is shown in Figure~\ref{rescheck} : as the resolution increases the algebraic decay shows negligible improvement beyond $N_x/L_x=4$. This suggests that the resolution chosen is sufficient to accurately capture the decay of the tails. The values of the exponent for $\bar{U}_x$ via the method exposed in section~\ref{S3} are tabulated in Table~\ref{expcheck}, with minimal changes in the computed values for resolutions higher than $N_x/L_x=4$.

The lower resolution $N_x/L_x=2$ features higher pointwise fluctuations of the low amplitude tails ($<10^{-5}$). This is primarily attributed to the difficulty of the spectral scheme to capture the singular (pointwise) nature of the initial condition (the Gibbs phenomenon \citep{canuto2007spectral}). These numerical artefacts were observed to decrease rapidly in amplitude with higher resolution, with a negligible impact on the value of the computed exponents. They have been observed in the results of Wf too (Spectral Fourier discretization) as well as in the pCf simulations performed using the spectral code SIMSON \citep{Duguet2010formation}.

The results presented in this study are all based on the domain size $L_x=L_z=1280$, with the resolution  $N_x,N_y,N_z=5120,33,10240$ for pCf, pPf and cPf, with $N_y=65$ for higher \emph{Re}. In the case of Wf the same domain has been used with a resolution of $N_x,N_y,N_z=4096,4,4096$.

\section{Decay exponents} \label{S3}

Whereas establishing that the decay of $\bar{U}_x$ and $\bar{U}_z$ is algebraic rather than exponential is straightforward once the numerical domain is large enough, an accurate and unambiguous numerical determination of the decay exponents is more difficult, mainly because the number of decades to assess the power-law behaviour is limited by computational limitations. Beyond the immediate yet inaccurate curve fitting  procedure, there is a well-established decomposition that makes the determination of these exponents more efficient, provided a few theoretical assumptions are made. \\

\subsection{Two-dimensional multipolar description} \label{S3.2}

\begin{figure}[h]
\centering
\includegraphics[scale=0.4]{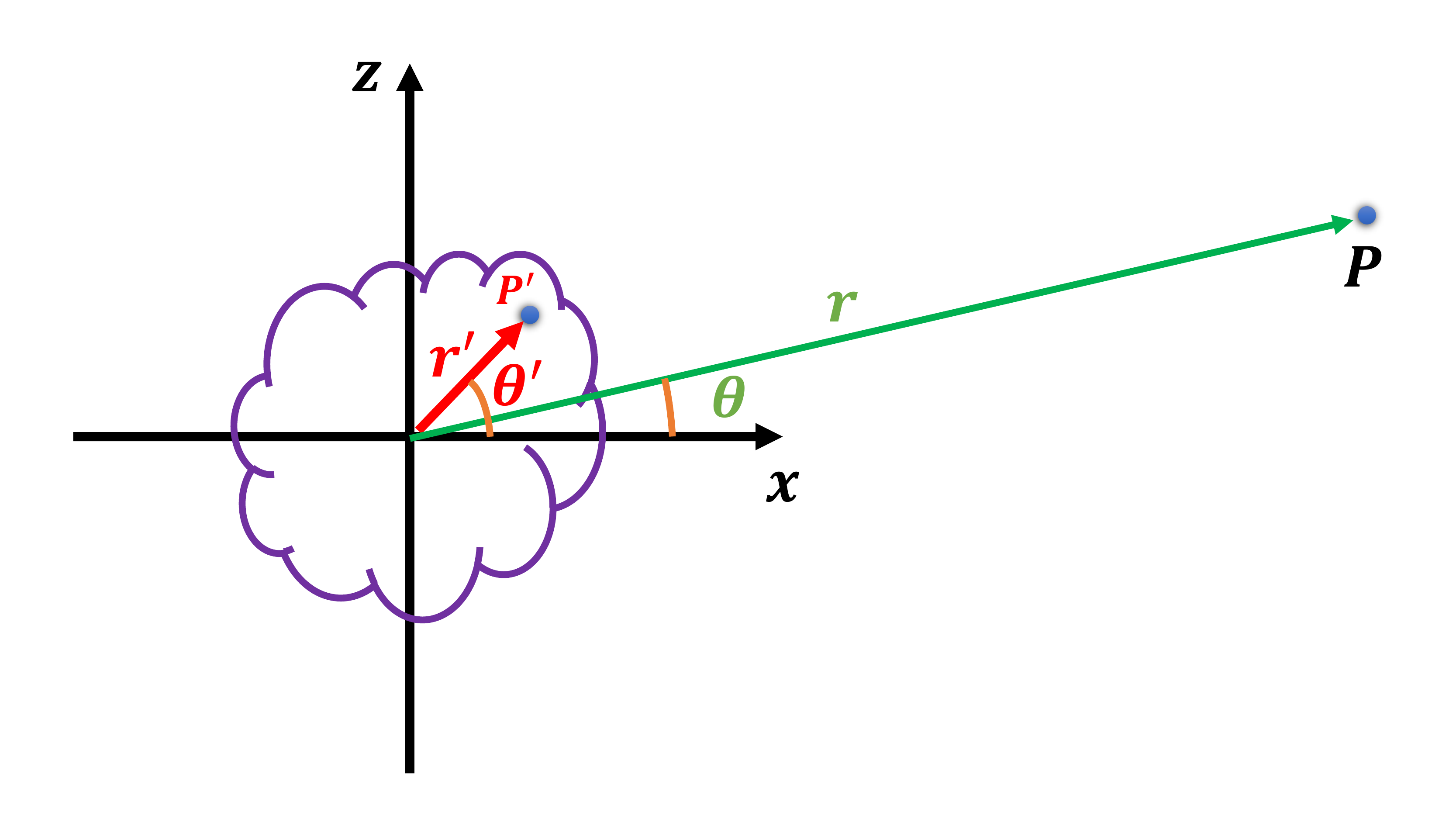}
\caption{Sketch of a concentrated vorticity source $\omega$ at the origin in a 2D plane}
\label{source}
\end{figure}

A quantitative analysis of the large-scale velocity field can be made using the $y$-averaged velocity field. While the structure of the large-scale flow for pCf (Fig~\ref{pcf_spot}) and Wf(Fig~\ref{wf_spot}) appears mainly \emph{quadrupolar}, that for both pPf (Fig~\ref{ppf_spot}) and cPf (Fig~\ref{cpf_spot}) is  reminiscent of a \emph{dipolar} flow. We proceed by analogy with the theory of two-dimensional incompressible flows, where the flow is described by a streamfunction $\psi$ and the velocity field derives from $\psi$ via $\bar{U}_x= \partial \psi/\partial z$ and $\bar{U}_z= - \partial \psi/\partial x$. We assume the existence of a vorticity field $\omega(x,z)$, oriented along ${\bm e_y}$, due to the presence of the spot. The $y$-averaged  field $\bm{\bar{U}}$ hence satisfies :
\begin{equation}
\nabla_{\perp} \cdot \bm{\bar{U}} = 0 \quad ; \quad \nabla_{\perp} \times \bm{\bar{U}} = \omega\bm{e_y},
\end{equation}
where $\nabla_{\perp}=\bm{e_x}\partial_x + \bm{e_z}\partial_z$ is the gradient operator with respect to the plane variables $x$ and $z$ only. The potential $\psi$ is linked to $\omega$ by the two-dimensional Poisson equation
\begin{equation}
\nabla_{\perp}^2 \psi = -\omega.
\end{equation} 
The generic solution of the Poisson equation above in unbounded two-dimensional space is
\begin{equation}
\psi = -\frac{1}{2 \pi} \int \ln\left|\bm{r}-\bm{r}^\prime \right| \, \omega (r^\prime,\theta^\prime) \, dS(r^\prime,\theta^\prime),
\label{green1}
\end{equation} 
where, $r=\sqrt{x^2+z^2}$ is the distance from the origin, $r^\prime=\sqrt{(x^\prime)^2+(z^\prime)^2}$ is the distance to the source from the origin, $dS$ is the area element. Eq. \ref{green1} can be linked to the sketch in figure \ref{source}. It expresses the contribution to $\psi$ at any point $P$ as a sum of the contributions from all points $P^\prime$ where the vorticity $\omega$ is non-vanishing.

For a concentrated distribution of vorticity of characteristic size $a$, located at the origin of the two-dimensional plane, the notion of far field corresponds to the hypothesis $r\gg a$ (Figure~\ref{source}). The multipole expansion of harmonic functions is a general solution of the  Stokes problem with sources distributed in a localised region only. It originates from the three-dimensional Biot-Savart integral formulation, after the denominator has been developed in powers of the small parameter $r'/r$ i.e. in the far-field where $r \gg r'$. Within this approximation, a Taylor expansion of the solution in Eq. (\ref{green1}) near the origin yields $\psi$ as a weighted sum
of multipoles, each order in the hierarchy decaying faster with $r$ than the previous one. In polar coordinates, this expansion writes formally
\begin{equation}
\psi(r,\theta) = -\frac{\ln(r)}{2 \pi} \, \int \omega \, dS + \sum_{k=1}^{\infty} \frac{1}{r^{k}} \int (r^\prime)^k \, \cos \left(k(\theta-\theta^\prime) \right) \, \omega \, dS,
\label{2me}
\end{equation}
In Eq. (\ref{2me}) the integration is carried over the whole plane, but because of the strong localisation of the vorticity (cf fig. \ref{Vy}) the main contribution to it is restrained to the area where the vorticity source $\omega$ is significantly non-zero.

The monopole corresponds in Eq. \ref{2me} to the prefactor $\ln(r)$
and its amplitude is directly proportional to the total vorticity. The dipole corresponds in Eq. (\ref{2me}) to the term with $k=1$, the quadrupole with $k=2$, etc... In the series for $\psi$ in Eq. (\ref{2me}), each term of order $k$ corresponds to a pole decaying as $r^{-k}$.

Although the present spot configuration is far from being two-dimensional, we postulate by analogy a multipolar expansion for the potential streamfunction $\psi$ from which $\bar{U}_x$ and $\bar{U}_z$, which are available directly from the simulations, are derived. We hypothesize that the expansion of Eq.~\ref{2me} is valid for the $y$-averaged velocity components  $\bar{U}_x$ and $\bar{U}_z$. 

The results from the previous section, namely the structure of the different flows, the two-dimensional incompressibility property, as well as the irrotational property outside the core of the spot, suggest that the two-dimensional description is relevant here for the effectively two-dimensional $y$-integrated velocity field. Importantly, as a consequence of the approximate curl-free property of the $y$-averaged flow $(\bar{U}_x,\bar{U}_z)$ (see Figure~\ref{Vy}), we can safely assume that monopole term is absent. 
In this situation only algebraic contributions remain in Eq. (\ref{2me}).
\begin{equation}
\psi(r,\theta) =  \sum_{k=1}^{\infty} \frac{1}{r^{k}} \int (r^\prime)^k \, \cos \left(k(\theta-\theta^\prime) \right) \, \omega \, dS,
\label{2mpole}
\end{equation}

\subsection{Results} \label{S3.3}

The multipolar expansion (\ref{2mpole}) allows one to separate the different algebraic contributions to the same flow field. This is of primary importance in an effort to identify the different multipoles and their respective algebraic decay exponent. The validity of the multipole expansion can be ascertained by decomposing the velocity field data in azimuthal Fourier harmonics at any radius $r$ in the domain and juxtaposing it with the numerical simulation data. The $k^{th}$ pole in the streamfunction is given by
\begin{equation}
\psi_k = \frac{1}{r^k} \int \, (r^\prime)^k \, \cos{(k(\theta-\theta^\prime))} \, \omega \, dS.
\end{equation}
The velocity components in polar coordinates are related to the streamfunction by:
\begin{equation}
\bar{U}_r = \frac{1}{r} \, \frac{\partial \psi}{\partial \theta}; \quad \bar{U}_\theta = -\frac{\partial \psi}{\partial r}
\end{equation}
\begin{align}
\bar{U}_r &= \frac{-k}{r^{k+1}} \int \, (r^\prime)^k \, \sin{(k(\theta-\theta^\prime))} \, \omega \, dS \\ 
\bar{U}_\theta &= \frac{k}{r^{k+1}} \int \, (r^\prime)^k \, \cos{(k(\theta-\theta^\prime)}) \, \omega \, dS 
\end{align}
The polar coordinate frame and the Cartesian frame of reference are related by
\begin{align}
{\bm e_r} &= \cos\theta \, {\bm e_x} \, \, + \, \, \sin\theta \, {\bm e_z}, \\
{\bm e_{\theta}} &= -\sin \theta \, \bm{e}_x \, \, + \, \, \cos\theta \, {\bm e_z}.
\end{align}
Converting the velocities from the polar frame of reference to the Cartesian frame, we find
\begin{align}
\bar{U}_x &= \frac{-k}{r^{k+1}} \, \int \, (r^\prime)^k \, \sin \left( (k+1)\theta-k\theta^\prime) \right) \, \omega \, dS \label{2dux} \\
\bar{U}_z &= \frac{k}{r^{k+1}} \, \int \, (r^\prime)^k \, \cos \left( (k+1)\theta-k\theta^\prime) \right) \, \omega \, dS. \label{2duz}
\end{align}
 
The azimuthal Fourier harmonics of the velocity components $\bar{U}_x$ and $\bar{U}_z$ hence display a signature for the $k^{th}$ pole at wavenumber $m=k+1$, rather than $m=k$. In particular the azimuthal Fourier amplitude spectra of $\bar{U}_x$ and $\bar{U}_z$ for the dipole velocity field ($k=1$) displays a peak at $m=2$,  while the quadrupolar velocity field ($k=2$) displays a peak at $m=3$. This property can be seen in e.g. Fig. 9 of Ref. \cite{lundbladh1991direct}.\\ 

The relative amplitude of the individual modes indicates the fraction of the energy content in that pole at distance $r$.  For a general velocity vector field containing all the poles, the relative energy content of a given pole $k$ we defined as
\begin{equation}
\hat{f}_k (r) = \frac{|A_k (r)|^2_{m=k+1}}{\sum\limits_{k=1}^{\infty} |A_k (r)|^2},
\end{equation}

where $A_k(r)$ is the ($r$-dependent) amplitude of the $k^{th}$ mode of the azimuthal Fourier spectrum of either $\bar{U}_x$ or $\bar{U}_z$ (shown in Eq~\ref{ak} with respect to $\bar{U}_x$).
\begin{equation}
    A_k (r) = \int_0^{2\pi} \bar{U}_x (r,\theta) \, \mathrm{e}^{i2\pi \theta \, k} \, d \theta \label{ak}
\end{equation}

\begin{figure}[h]
	\centering
	\includegraphics[scale=0.32]{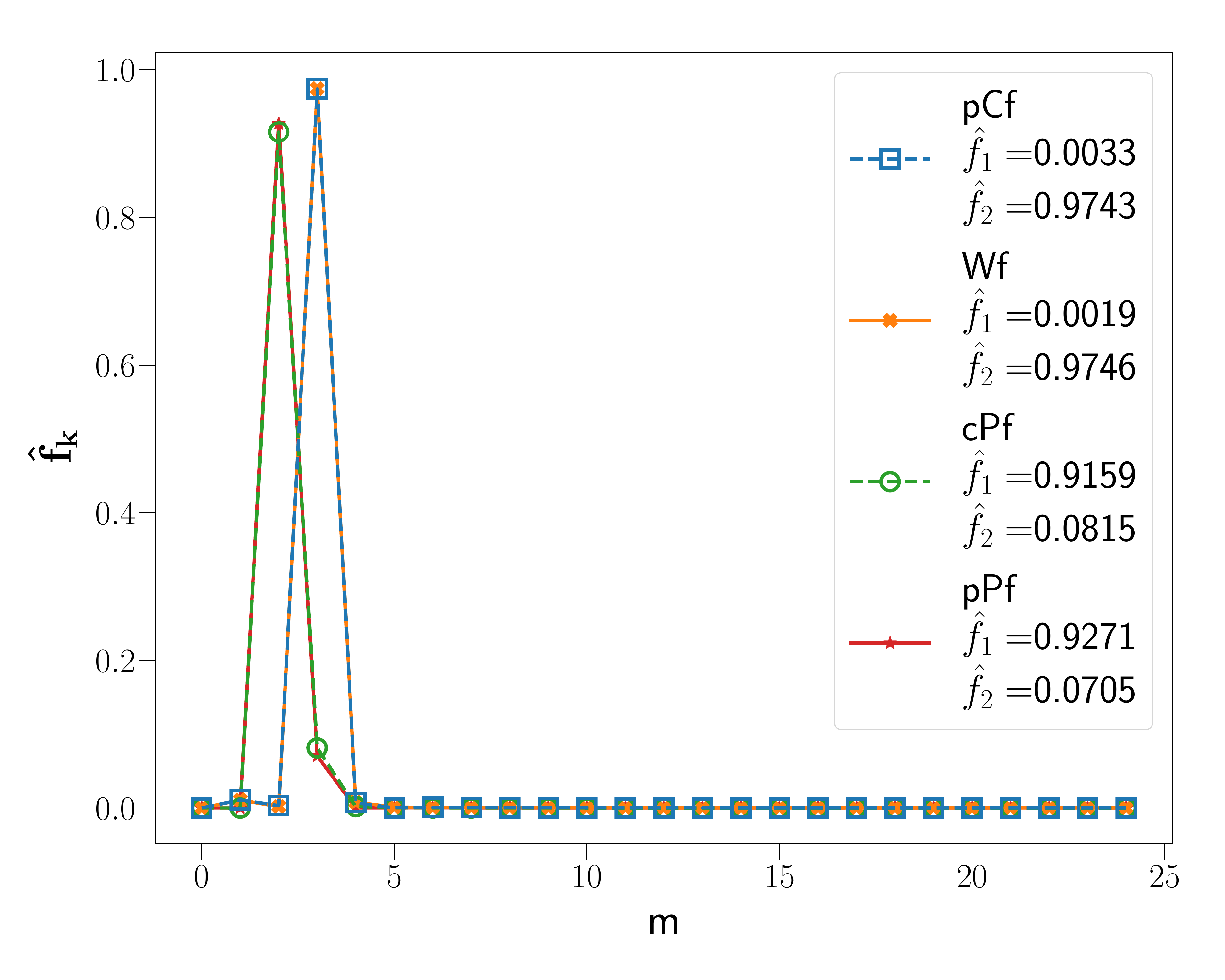}
	\caption{Variation of $\hat{f}_k$ for the $y$-averaged velocity component $\bar{U}_z$ at radius $r_o=400$ depicted for pCf at $Re=400$ and $T=300$, Wf at $Re_V=83.5$ and $T=250$, cPf at $Re_W=700$ and $T=200$ and pPf at $Re_{cl}=3200$ and $T=200$}
	\label{qlsf}
\end{figure}
The quantification of the relative energy contents from the energy spectra is shown at a specified radius $r=400$ for all the different flow scenarios in Figure~\ref{qlsf}. Both pCf and Wf verify $\hat{f}_2>0.9$. This suggests that most of their energy at this radius is contained in the $m$=3 mode, corresponding to the quadrupole in the expansion in Eq. (\ref{2mpole}). The case of pPf and CPf is different : the energy is mostly contained in the mode $m$=2 corresponding to the dipole, as attested by $\hat{f}_1>0.9$. This is completely consistent with the qualitative conclusions drawn visually from figure \ref{spots}. 

\begin{figure}[h]
     \centering
     \begin{subfigure}[b]{0.48\textwidth}
         \centering
         \includegraphics[width=\textwidth]{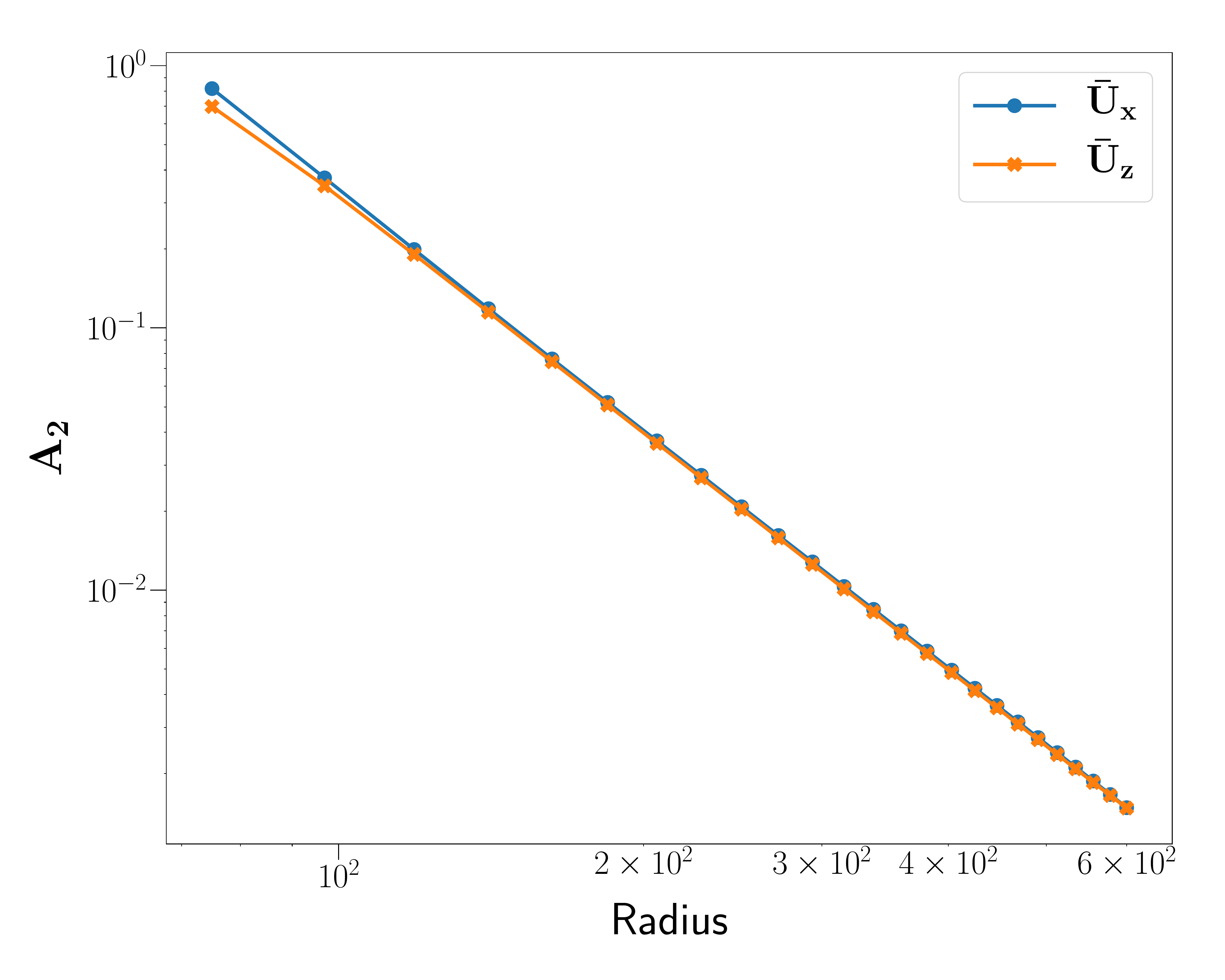}
         \caption{}
         \label{pcf_tail}
     \end{subfigure}
     \hfill
     \begin{subfigure}[b]{0.48\textwidth}
         \centering
         \includegraphics[width=\textwidth]{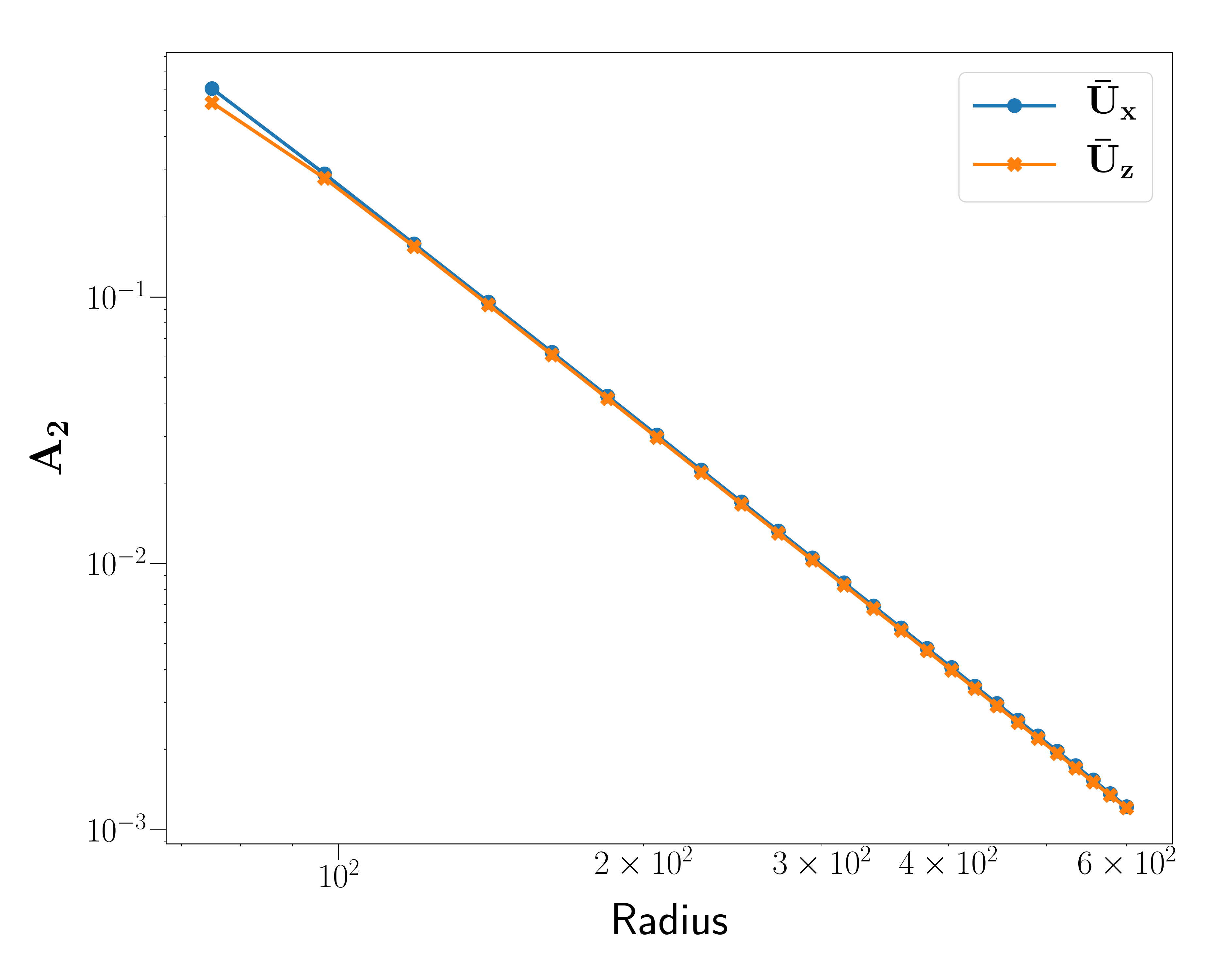}
         \caption{}
         \label{wf_tail}
     \end{subfigure}
     \hfill
     \begin{subfigure}[b]{0.48\textwidth}
         \centering
         \includegraphics[width=\textwidth]{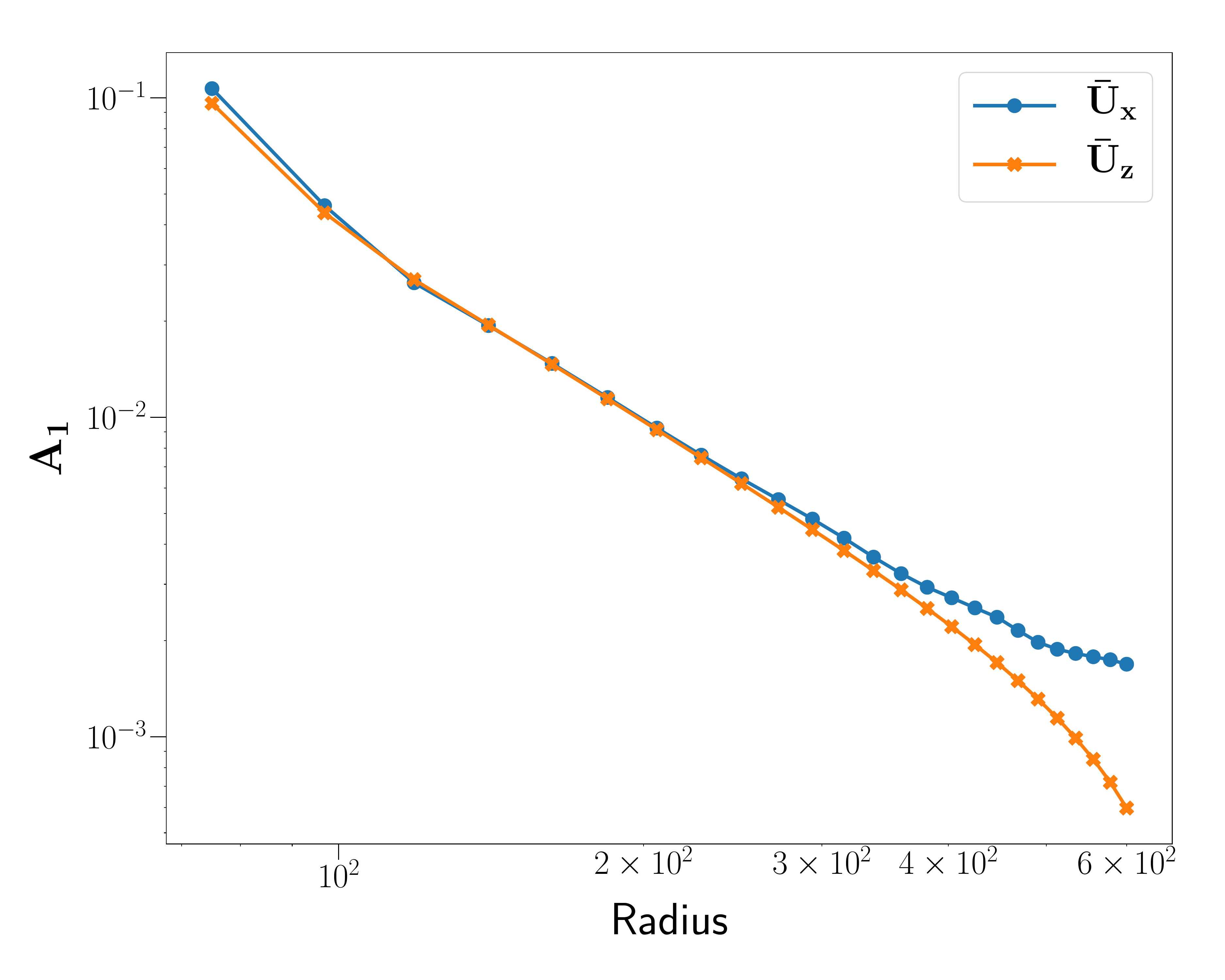}
         \caption{}
         \label{ppf_tail}
     \end{subfigure}
     \hfill
     \begin{subfigure}[b]{0.48\textwidth}
         \centering
         \includegraphics[width=\textwidth]{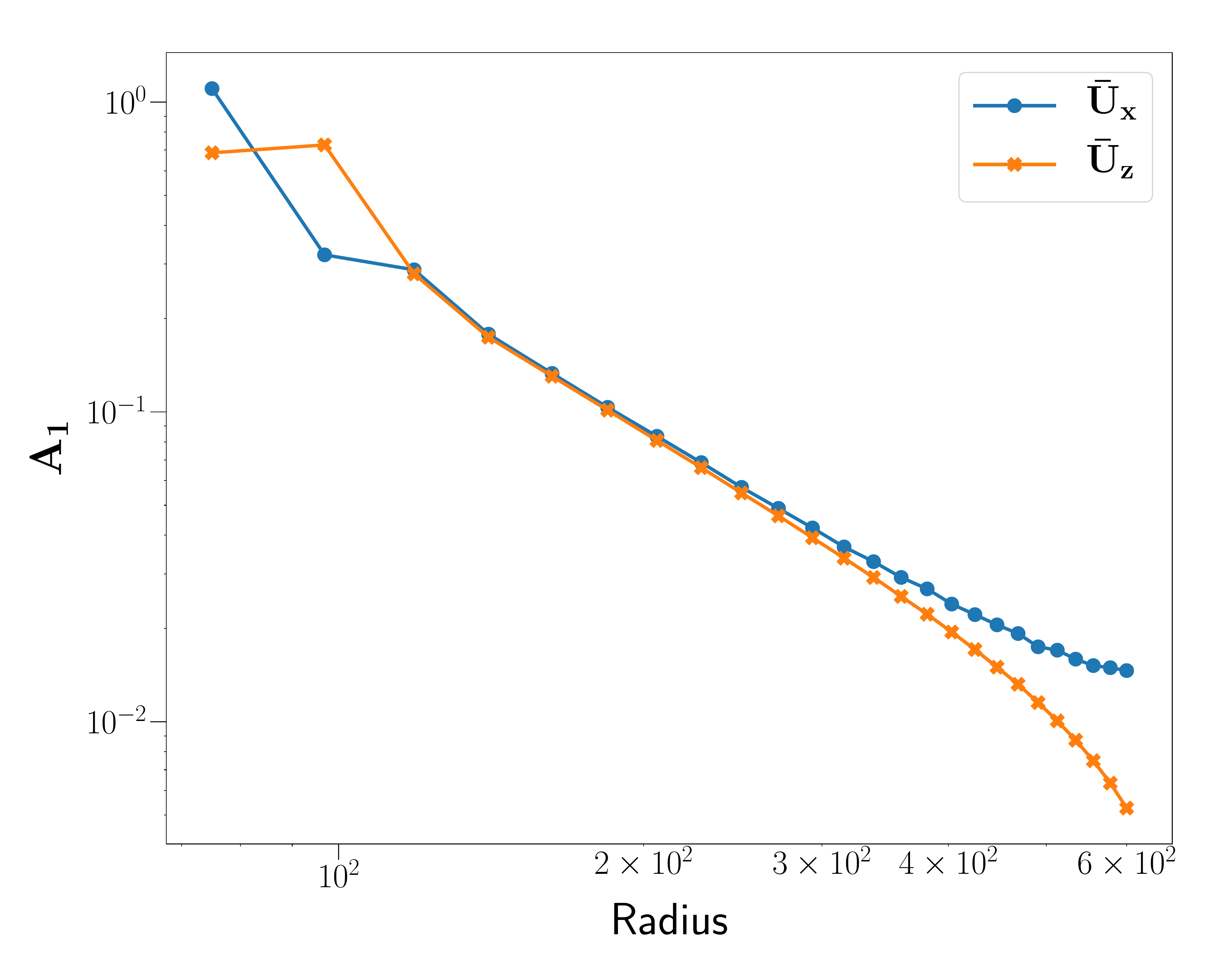}
         \caption{}
         \label{cpf_tail}
     \end{subfigure}
        \caption{Log-Log plot of the amplitude of the dominant modes $m=2$ for dipole and $m=3$ for quadrupole against the radius $r$ for both $\bar{U}_x$ and $\bar{U}_z$  (a) pCf at $Re=400$ and $T=300$ (b) Wf at $Re_V=83.5$ and $T=250$ (b) CPf at $Re_W=700$ and $T=200$ (d) pPf at $Re_{cl}=3200$ and $T=200$}
        \label{tail_decay}
\end{figure}

The radial decay is analysed by plotting the amplitude of the dominant mode, extracted from the spectrum, as a function of $r$ in Figure~\ref{tail_decay}. All plots display a range of almost a decade where the decay is well fitted by a power-law. The decay exponent $\alpha$ is computed classically as
\begin{equation}
\alpha = - \left< \frac{\ln (A_k (r)/A_k (r_1))}{\ln(r/r_1)}\right>\\
\label{eqn}
\end{equation}
In Eq. \ref{eqn}, radial values $r$ are sampled evenly in $(75,600)$ and $\left< \, \cdot \, \right>$ indicates its average value. \\

\begin{minipage}{\textwidth}
  \begin{minipage}[b]{0.48\textwidth}
    \centering
    \begin{tabular}{|c|c|c|}
    	\hline
	    \multicolumn{3}{|c|}{pCf} \\ \hline
        $Re$  & $\alpha$ ($\bar{U}_x$) & $\alpha$ ($\bar{U}_z$) \\ \hline
        400 & $+3.02 \pm 0.01$ & $+3.01 \pm 0.01$ \\ \hline
        500 & $+3.02 \pm 0.02$ & $+3.01 \pm 0.01$ \\ \hline
        600 & $+2.97 \pm 0.16$ & $+2.98 \pm 0.04$ \\ \hline
    \end{tabular}
    \captionof{table}{Decay exponents for the algebraic decay of $\bar{U}_x$ and $\bar{U}_z$ in pCf for different \emph{Re} numbers}
	\label{pcf_exp}
  \end{minipage}
  \hfill
  \begin{minipage}[b]{0.48\textwidth}
    \centering
    \begin{tabular}{|c|c|c|}
    	\hline
    	\multicolumn{3}{|c|}{Wf} \\ \hline
        $Re_V$  & $\alpha$ ($\bar{U}_x$) & $\alpha$ ($\bar{U}_z$) \\ \hline
        83.5 & $+2.99 \pm 0.01$ & $+2.97 \pm 0.02$ \\ \hline
        104.5 & $+3.01 \pm 0.01$ & $+2.93 \pm 0.05$ \\ \hline
        125.5 & $+3.01 \pm 0.01$ & $+2.91 \pm 0.08$ \\ \hline
    \end{tabular}
    \captionof{table}{Decay exponents for the algebraic decay of $\bar{U}_x$ and $\bar{U}_z$ in Wf for different \emph{Re} numbers}
	\label{wf_exp}
  \end{minipage}
  \hfill
  \begin{minipage}[b]{0.48\textwidth}
    \centering
    \begin{tabular}{|c|c|c|}
    	\hline
	    \multicolumn{3}{|c|}{pPf} \\ \hline
        $Re_{cl}$  & $\alpha$ ($\bar{U}_x$) & $\alpha$ ($\bar{U}_z$) \\ \hline
        1800 & $+1.94 \pm 0.01$ & $+2.12 \pm 0.09$ \\ \hline
        3200 & $+1.98 \pm 0.1$ & $+2.11 \pm 0.09$ \\ \hline
        5000 & $+2.11 \pm 0.3$ & $+2.11 \pm 0.08$ \\ \hline
    \end{tabular}
    \captionof{table}{Decay exponents for the algebraic decay of $\bar{U}_x$ and $\bar{U}_z$ in pPf for different \emph{Re} numbers}
	\label{ppf_exp}
  \end{minipage}
  \hfill
  \begin{minipage}[b]{0.48\textwidth}
    \centering
    \begin{tabular}{|c|c|c|}
    	\hline
	    \multicolumn{3}{|c|}{cPf} \\ \hline
        $Re_W$  & $\alpha$ ($\bar{U}_x$) & $\alpha$ ($\bar{U}_z$) \\ \hline
        600 & $+1.83 \pm 0.05$ & $+2.06 \pm 0.1$ \\ \hline
        700 & $+1.88 \pm 0.07$ & $+2.08 \pm 0.1$ \\ \hline
        800 & $+1.87 \pm 0.06$ & $+2.08 \pm 0.1$ \\ \hline
    \end{tabular}
    \captionof{table}{Decay exponents for the algebraic decay of $\bar{U}_x$ and $\bar{U}_z$ in cPf for different \emph{Re} numbers}
	\label{cpf_exp}
  \end{minipage}
\end{minipage}
\vspace*{1cm}

The far field decay of ${\bm {\bar{U}}}$ for both pCf and Wf is governed by a decay exponent +3 (cf Table~\ref{pcf_exp},\ref{wf_exp}), as predicted in two dimensions (Eq. \ref{2dux} and \ref{2duz}) for a quadrupolar flow. The accuracy on the exponent is better than one percent. For pPf and cPf however, the exponent can be approximated as +2 within less than 6\% (cf Table~\ref{ppf_exp},\ref{cpf_exp}), as expected in two dimensions for a dipolar flow.  We emphasize that these decay exponents +2 (dipole) and +3 (quadrupole) concern the velocity components $\bar{U}_x$ (Eq.~\ref{2dux}) and $\bar{U}_z$ (Eq.~\ref{2duz}), not the streamfunction as in classical three-dimensional hydrodynamics. These tables show that these exponents, as well as the qualitative structure of these flows, are independent of \emph{Re}.  In all scenarios we note a tendency towards lesser accuracy as $Re$ increases, which can be attributed to the faster growth of the turbulent spot and hence the stronger finite-size effects at a finite time.\\

\begin{figure}[h]
    \centering
	\includegraphics[scale=0.3]{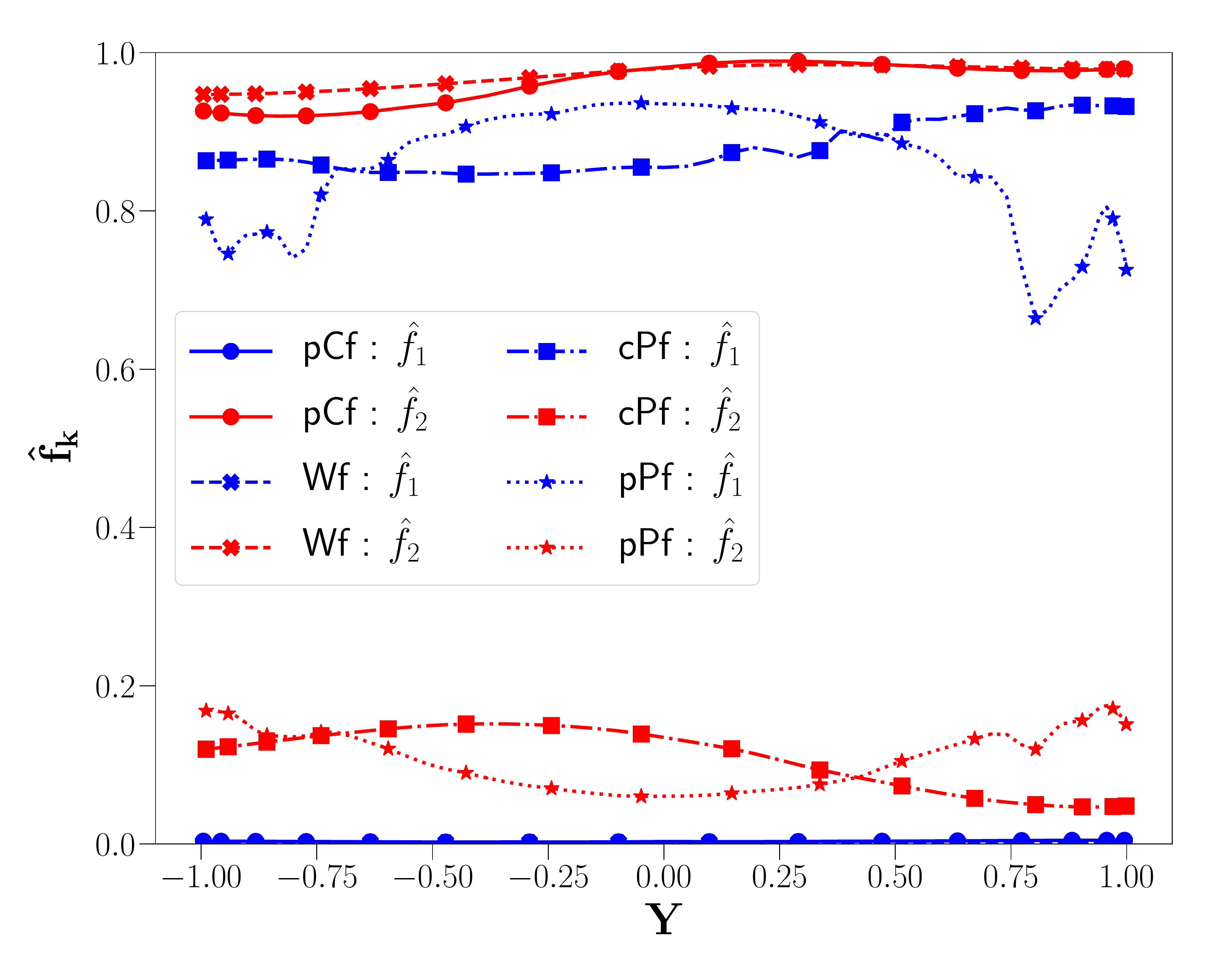}
	\caption{Variation of the fraction of energy content in the dipole : $\hat{f}_1$ (blue) and quadrupole : $\hat{f}_2$ (red) as a function of the wall normal coordinate at radius of $r_o=300$ from the origin for different flows. pCf at $Re=400$ and $T=300$ ; Wf at $Re_V=83.5$ and $T=250$ ; cPf at $Re_W=700$ and $T=200$ ; pPf at $Re_{cl}=3200$ and $T=200$}
	\label{tails_y}
\end{figure}

While $y$-averaging captures well the large-scale features of the velocity field, the actual flow field is inherently three-dimensional. Further knowledge about the dependence of the real velocity field on the wall-normal coordinate $y$ is hence necessary. 
The fraction of energy content in the dipole and quadrupole have been monitored for the different flow scenarios as a function of $y$, and the results are shown in Figure~\ref{tails_y}. The global trend is that these energy contents are weakly dependent on $y$. For pCf and Wf the quadrupole remains dominant with $\hat{f}_2>0.9$ and negligible energy content in the dipole $\hat{f}_1<0.02$. In pPf and cPf the dominant term is still the dipole with $\hat{f}_1>0.8$ while for the quadrupole $\hat{f}_1<0.2$. Fig. \ref{tails_y} suggests that the quadrupolar component of CPf is slightly enhanced
near the middle of the gap compared to pPf, however the levels of $\hat{f}_1$ and  
$\hat{f}_2$ remain very comparable between these two flows. This suggests that spots in CPf
should not be interpreted as a linear superposition of contributions from pPf and pCf.

\section{Discussion } \label{S4}

Two classes of wall-bounded shear flows emerge from our analysis : dipole-like fields for which the planar velocity field decay is dominated by $O(r^{-2})$ and quadrupole-like fields for which it decays like $O(r^{-3})$, with $r$ the distance from the origin of the turbulent spot. These findings rule out exponential decay, which is re-interpreted as a finite-size effect. The diversity of the flow scenarios and  the independence of the results to varying numerical resolution and domain size highlight their robustness. Beyond these observations, several comments can be made.

\subsection{Symmetries}

The multipolar expansion of Eq. (\ref{2mpole}) used in Section 3 assumes that, to leading order, two flow components dominate the far field decay : a dipole and a quadrupole. The dipole shows a slower decay than the quadrupole and is expected to dominate in the far field, except if its amplitude vanishes. Monopoles are ruled out by the irrotationality of the $y$-integrated velocity field, while further orders are associated with negligible contributions, and would decay faster anyway. 
Whether spots for a given flow case will be dominated by a dipole or a quadrupole is determined by the symmetries of the flow system only. Both pCf and Wf satisfy the rotational symmetry $S$ defined by
\begin{align}
S \quad &: \quad [U_x,U_y,U_z](x,y,z) \xrightarrow{} [-U_x,-U_y,U_z](-x,-y,z). \label{sz}
\end{align}
Although $S$ is not imposed in the computations, it is satisfied exactly by the laminar flow solutions of pCf and Wf and their turbulent mean flow.  By virtue of $y$-averaging, this symmetry of the 3D flow is associated with the  $S_2$ symmetry for the two-dimensional flow ${\bm {\bar{U}}}$, defined by 
\begin{align}
S_2 \quad &: \quad [U_x,U_z](x,z) \xrightarrow{} [-U_x,U_z](-x,z). \label{sz2}
\end{align}
When $S$ is exactly satisfied, $S_2$ is satisfied too, and the dipolar component $\hat{f_{1}}$ cancels out exactly as in pCf and Wf. The decay exponent for ${\bm \bar{U}}$ is then +3 because the next leading order term is the quadrupole. The dipole, when present as in CPf and pPf, is found to orient itself according to the imposed pressure gradient as evidenced in Fig~\ref{cpf_spot} and Fig~\ref{ppf_spot}. On the contrary if $S$ is not satisfied, the dipolar component $\hat{f_{1}}$ does not vanish and necessarily dominates the decay in the far field, as seen in pPf and CPf. 
The flow fields that satisfy the symmetry $S$ are those permitting quadrupolar spots, and correspond here to the flow cases with no pressure gradient applied.

\subsection{Analogies with confined active matter} \label{S4.2}

We recall that the algebraic decay exponents +2 and +3 identified for, respectively, the dipolar and the quadrupolar flow components of turbulent spots do not match the textbooks predictions for unbounded three-dimensional hydrodynamics, where these exponents are respectively +3 and +4 for velocity fields \cite{alekseenko_theory_2007,KIM199113}. However these exponents do match the two-dimensional theory of incompressible flows induced by concentrated vorticity sources \citep{batchelor_2000}. A full derivation based on the Navier-Stokes equations is cumbersome and leads to such results only at the price of strong analytical simplifications, see Ref. \cite{zhe2020} for an illustration. Instead we note that similar analytical results have been put forward recently in the context of microfluidics applied to active matter.
The  collective motion of living microscopic organisms such as bacteria or plankton is governed by low Reynolds number hydrodynamics, namely by the Stokes regime with $Re \ll$1. It is common practice to focus on the flow field induced by a single such micro-organism in isolation before generalising to collective interactions. In unbounded flows, a pointwise swimmer is usually described as a steady Dirac-like forcing term to both the continuity and the momentum equations, both equations being linear in the low-$Re$ regime. This gives rise to a distinction between so-called pushers and pullers depending on the force exerted on the fluid by the organism. In the context of a confined flow, e.g. inside a Hele--Shaw cell, recent studies have demonstrated that the situation is different \cite{diamant_hydrodynamic_2009,spagnolie_hydrodynamics_2012}. Confinement leads to screening of the dominant terms and to a re-ordering of the dominant far field contributions \cite{brotto_hydrodynamics_2013}. The leading order term turns out to be a source dipole with a velocity field decaying like $O(r^{-2})$ just like in our analysis. This has lead some authors to suggest an effective two-dimensional model \cite{jeanneret_confinement_2019}, whose derivation from first principles is not exact.

In the present study of turbulent spots, the large-scale flow in the far field is triggered by the flow itself, not by an external obstacle. No force singularity is required in this context, however  decomposing the flow field as a sum of vortex source singularities is still relevant. By analogy with the confined Stokes regime, we have here suggested a similar decomposition of the two-dimensional field ${\bm \bar{U}}$. 

The results comfortably confirm that the effective two-dimensional picture also works for localised turbulent structures at finite Reynolds numbers and in the presence of a mean shear, a framework relatively far from the original microfluidics context. The role of the shear for turbulent spots is not entirely understood yet, however it should not be underestimated: shear, via complex nonlinear mechanisms, sustains the strongest perturbations against viscous decay \cite{waleffe1997self}.

\subsection{Collective effects} \label{S4.3}

In the context of microfluidics, the singular solutions of the source or force type are considered as Green's functions associated with the Stokes problem : linear responses to pointwise forcing \cite{pozrikidis1997introduction}. The Green's formalism has the advantage that, out of the knowledge of one localised response, it is possible to reconstruct the flow field corresponding to a set of multiple sources. Despite the nonlinear context, the present findings about the far field associated with one turbulent spot suggest possible generalisations. For instance turbulent patches larger than a single spot can be modelled as the superimposition of several spots, here of equal strength yet with different spatial origins. In particular, it would be interesting to see how much oblique turbulent stripes (whose far field decay is not algebraic) can be expressed as an array of individual turbulent spots,
and whether interaction distances between stripes can be deduced from the screening effects \cite{gome2020statistical}.
Such aspects have not been investigated here and are left for future studies.

\section{Conclusion} \label{S5}

Direct numerical simulations of turbulent spots have been performed in large numerical domains in different wall-bounded shear flows such as plane Couette flow (pCf), plane Poiseuille flow (pPf), Couette-Poiseuille (CPf) and the Waleffe flow (Wf). The large-scale flows generated by these spots were found to be of either quadrupolar nature for pCf and Wf or of dipolar nature for pPf and CPf. This difference in behavior is attributed to the different symmetries of the laminar flows, which depend in turn on the presence or not of an applied pressure gradient. The velocity tails were found to decay algebraically in all scenarios, with a decay exponent of $\approx +3$ for the quadrupole and $\approx +2$ for the dipole. The exponents have been also found to be independent of \emph{Re}. This suggests to revisit numerically or analytically some of the works where exponential scalings have been claimed, e.g. Ref. \cite{Brand2014doubly}, in larger domains. These results are consistent with a two-dimensional description of the flow based on a localised vorticity source and with the associated multipolar decomposition. The present results are also consistent with active matter theories in confined geometries. Despite a different range of parameters, the dominant multipolar components found around turbulent spots look indistinguishable from the source singularities put forward in confined Stokes flows. 

\section{Acknowledgements}

Y. D. and M. C. would like to acknowledge the late Bruno Eckhardt for his continuous encouragements to perform this study. P.K. and Y.D. would like to acknowledge and thank the entire team of \emph{channelflow.ch} for building the code and making it open source. This study was made possible by the grant of computational resources from IDRIS (Institut du D\'eveloppement et des Ressources en Informatique Scientifique) and the support of its staff in helping us efficiently utilize those resources. The authors would also like to thank Jalel Chergui, Florian Reetz, Laurette S. Tuckerman, Sebastian Gom\'e,  Romain Monchaux and Philipp Schlatter for their valuable discussions and technical inputs.

\bibliography{biblio_final.bib}

\end{document}